\documentclass[%
 reprint,
 aps,
]{revtex4-2}
\usepackage{xcolor}
\usepackage{graphicx}
\usepackage{dcolumn}
\usepackage{bm}
\usepackage{amsmath}
\usepackage{comment}

\usepackage[]{lineno}

\begin{document}

\title{Contrastive learning through implicit non-equilibrium memory}
\title{Temporal Contrastive Learning: contrastive learning through implicit non-equilibrium memory}
\title{Temporal Contrastive Learning through implicit non-equilibrium memory}
\author{Martin J. Falk${}^{1,+}$, Adam T. Strupp${}^{1,+}$, Benjamin Scellier${}^2$, Arvind Murugan${}^1$}
\affiliation{${}^+$denotes equal contribution}
\affiliation{${}^1$Department of Physics, University of Chicago, Chicago, IL 60637}
\affiliation{${}^2$Rain AI, San Francisco, CA 94110}

\date{\today}

\begin{abstract}
The backpropagation method has enabled transformative uses of neural networks. 
Alternatively, for energy-based models, local learning methods involving only nearby neurons offer benefits in terms of decentralized training, and allow for the possibility of learning in computationally-constrained substrates.
One class of local learning methods \textit{contrasts} the desired, clamped behavior with spontaneous, free behavior.
However, directly contrasting free and clamped behaviors requires explicit memory. 
Here, we introduce `Temporal Contrastive Learning', an approach that uses integral feedback in each learning degree of freedom to provide a simple form of implicit non-equilibrium memory. 
During training, free and clamped behaviors are shown in a sawtooth-like protocol over time.
When combined with integral feedback dynamics, these alternating temporal protocols generate an implicit memory necessary for comparing free and clamped behaviors, broadening the range of physical and biological systems capable of contrastive learning. 
Finally, we show that non-equilibrium dissipation improves learning quality and determine a Landauer-like energy cost of contrastive learning through physical dynamics. 
\end{abstract}

                              
\maketitle


\color{black}
The modern success of neural networks is underpinned by the backpropagation algorithm, which easily computes gradients of cost functions on GPUs\cite{krizhevsky2012imagenet}.
Backpropagation is a `non-local' operation, requiring a central processor to coordinate changes to a synapse that can depend on the state of neurons far away from the synapse.  While powerful, such methods may not be available in physical or biological systems with strong constraints on computation and communication. 

\begin{figure}
\includegraphics[width=\linewidth]{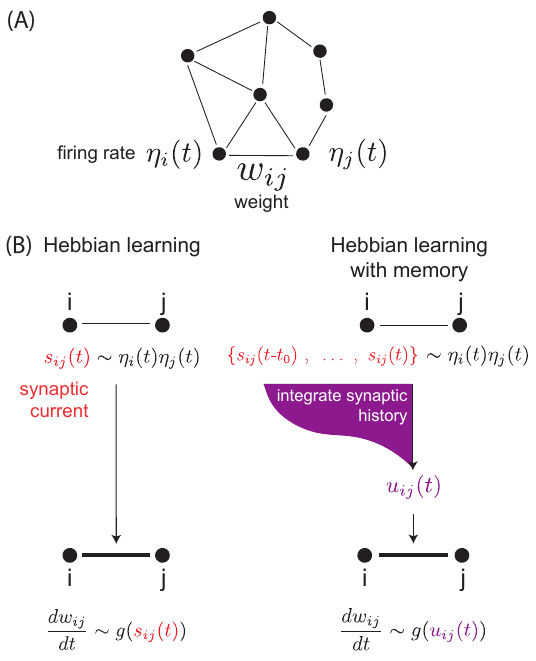}
\caption{\textbf{Hebbian learning with memory.} (A) Consider a neural network with neuron $i$ firing at rate $\eta_i(t)$ and synaptic weight $w_{ij}$ between neurons $i,j$. (B) In Hebbian learning, the weights $w_{ij}$ are changed as a function $g$ of the synaptic current $s_{ij}(t) \sim \eta_i(t) \eta_j(t)$. Here, we generalize the Hebbian framework to a model in which weights $w_{ij}$ are changed based on the history of $s_{ij}(t)$, i.e., based on $u_{ij}(t) = \int_{-\infty}^{t} K(t-t') s_{ij}(t')dt'$ where $K$ is a memory kernel. We find that non-monotonic kernels $K$ that arise in non-equilibrium systems encode memory that naturally enables contrastive learning through local rules.
}
\label{fig:1schematic}
\end{figure}

A distinct thread of learning theory has sought  `local' learning rules, such as the Hebbian rule (`fire together, wire together'), where updates to a synapse are based only on the state of adjacent neurons (Fig. \ref{fig:1schematic}). Such local rules allow for, e.g. distributed neuromorphic computation\cite{grollier2020neuromorphic,burr2017neuromorphic,shastri2021photonics}. Excitingly, local learning rules also open the possibility of endowing computationally-constrained 
physical systems with functionality through an \textit{in situ} period of training, rather than by prior backpropagation-aided design on a computer\cite{stern2023learning,murugan2015multifarious, zhong2017associative,evans2022pattern,poole2017chemical,poole2023autonomous,pashine2019directed,stern2020continual,stern2020supervised,stern2021supervised,anisetti2023learning,anisetti2024frequency,patil2023self,arinze2023learning,altman2023experimental,behera2023enhanced,wang2024training}.
In these settings, sometimes as simple as chemical reactions within a cell or a mechanical material, there is no centralized control that would allow backpropagation to be a viable method of learning. Consequently, local rules allow for the intriguing possibility of autonomous `physical learning'\cite{stern2023learning, stern2021supervised} --- no computers or electronics needed --- in a range of both natural and artificial systems of constrained complexity such as molecular and mechanical networks. 

In particular, a large class of local `contrastive learning' algorithms (contrastive Hebbian learning\cite{movellan1991contrastive, xie2003equivalence,baldi1991}, Contrastive Divergence\cite{hinton2002training}, Equilibrium Propagation\cite{scellier2017equilibrium})  promise impressive results, but make requirements on the capabilities of a single synapse (or more generally, on learning degrees of freedom). While details differ, training weights generally are updated based on the difference between Hebbian-like rules applied during a `clamped' state that roughly corresponds to desired behaviors and a `free' state that corresponds to the spontaneous (and initially undesirable) behaviors of the system. 

Therefore, a central obstacle for autonomous physical systems to exploit contrastive learning is that weight updates require a comparison between free and clamped states, but these states occur at different moments in time. 
Such a comparison requires memory at each synapse to store free and clamped state information in addition to global signals that switch between these free and clamped memory units and then retrieve information from them to perform weight updates\cite{yi2023activity,laydevant2024training}. These requirements make it difficult to see how contrastive learning can arise in natural physical and biological systems and demand additional complexity in engineered neuromorphic platforms. 

Here, our primary contribution is to show how a ubiquitous process - integral feedback control - can allow for contrastive learning without the complexities of explicit memory of free and clamped states or switching between Hebbian and anti-Hebbian update modes. Further, in our method, computing the contrastive weight update signal and then performing the update of weights are not separate steps involving different kinds of hardware but are the same unified \textit{in situ} operation. This approach to contrastive learning, which we call `Temporal Contrastive Learning through feedback control', allows a wide range of physical and biological platforms without central processors to `physically learn' (i.e., autonomously learn) novel functions through contrastive learning methods.

\color{black}
First, we introduce a simple model of implicit memory in integral feedback-based update dynamics at synapses. Synaptic weights $w_{ij}$ are effectively updated by $\int^t_{-\infty} K(t-t') s_{ij}(t') dt'$, where $K$ is a memory kernel. Updates of this form arise naturally due to integral feedback control\cite{doyle2013feedback} of synaptic current $s_{ij}$, without the need for an explicit memory element. 


Using a microscopically reversible model of the dissipative dynamics at each synapse, we are able to calculate an energy dissipation cost of contrastive learning. The dissipation can be interpreted as the cost of (implicitly) storing and erasing information about free and clamped states over repeated cycles of learning. 

Finally, we propose how  the non-equilibrium memory needed for contrastive learning arises naturally as a by-product of integral feedback control\cite{doyle2013feedback, yi2000robust,lan2012energy} in many different physical systems. As a consequence, a wide range of physical and biological systems might have a latent ability to learn through contrastive schemes by exploiting feedback dynamics.

\color{black}
\section*{Background: Contrastive Learning}

Contrastive Learning (CL) was introduced in the context of training Boltzmann machines\cite{ackley1985learning} and has been developed further in numerous works\cite{movellan1991contrastive,xie2003equivalence,baldi1991,hinton2002training,scellier2017equilibrium}. 

\begin{figure*}
\includegraphics[width=\linewidth]{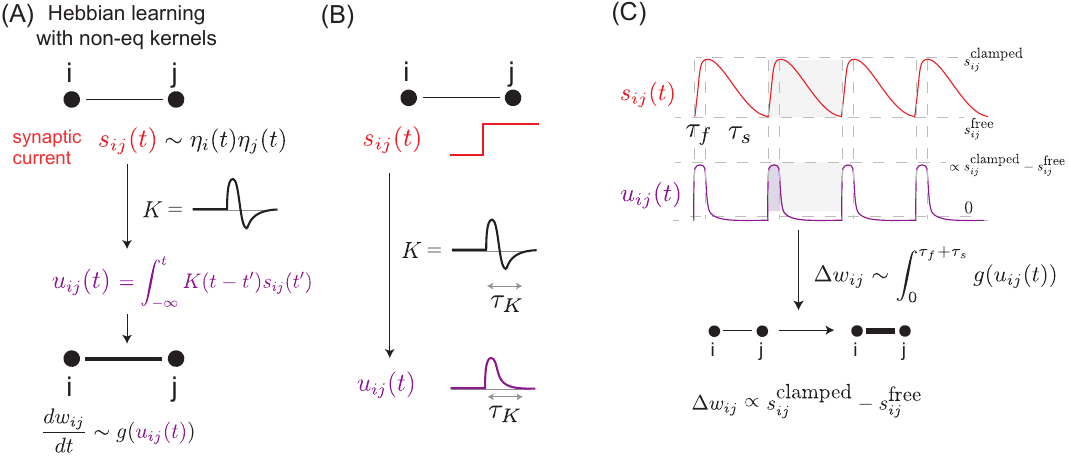}
\caption{\textbf{Contrastive learning through non-monotonic memory kernels} 
(A) We consider a network whose weights $w_{ij}$ are updated by a history of the synaptic current $s_{ij}(t)$, i.e., by $u_{ij}(t) = \int_{-\infty}^{t} K(t-t') s_{ij}(t')dt'$ with a non-monotonic kernel $K$ of timescale $\tau_K$ as shown. $g(x)$ is a non-linear function of the type in Eq. \ref{eqn:nonlinearity}. 
(B) Response of $u_{ij}$ to a step change in $s_{ij}$ shows that $u_{ij}$ effectively computes a finite-timescale derivative of $s_{ij}(t)$; $u_{ij}$ responds to fast changes in $s_{ij}$ but is insensitive to slow or constant values of $s_{ij}(t)$. (C) Training protocol: Free and clamped states are seen in sequence, with the free state rapidly followed by the clamped state (fast timescale $\tau_f$) and clamped boundary conditions slowly relaxing back to free (slow timescale $\tau_s$). The rapid rise results in a large $u_{ij}(t)$ whose integral reflects the desired contrastive signal; but the slow fall results in a small $u_{ij}(t)$ whose impact on $w_{ij}$ is negligible, given the nonlinearity $g(u_{ij}(t))$. Thus, over one sawtooth period, weights updates $\Delta w_{ij}$ are proportional to the contrastive signal $s_{ij}^{\rm clamped} - s_{ij}^{\rm free}$. }
\label{fig:1schematic-pt2}
\end{figure*}

While CL was originally introduced and subsequently developed for stochastic systems (such as Boltzmann machines) \cite{ackley1985learning,hinton2002training,agoritsas2023explaining}, here we review the version of \cite{movellan1991contrastive,scellier2017equilibrium} for deterministic systems (such as Hopfield networks). CL applies in systems described by an energy function $E$ (more accurately, a Lyapunov function). The system may be supplied with a boundary input and evolves towards a minimum of $E$, called `free state'. The system may also be supplied with a boundary desired output (in addition to the boundary input), driving the system towards a new energy minimum, called `clamped state'. Contrastive training requires updating weights as:
\begin{equation}
\Delta w_{ij} = \epsilon(s_{ij}^{\rm clamped} - s_{ij}^{\rm free}),
\label{eqn:wijfreeclamped}
\end{equation}
where $\epsilon$ is a learning rate and $s_{ij}^{\rm free},s_{ij}^{\rm clamped}$ are the synaptic currents at the free and clamped states, defined by $s_{ij} = \frac{\partial E}{\partial w_{ij}}$. For example, for a neural network with a Hopfield-like energy function, we recover a conventional Hebbian rule with $s_{ij} = \eta_i \eta_j$, where $\eta_i$ is the firing rate of neuron $i$ (Fig. \ref{fig:1schematic}B). 
\color{black}

A key benefit of contrastive training is that it is a \emph{local} rule; synaptic weight $w_{ij}$ is only updated based on the state of neighboring neurons $i,j$. Hence, it has been proposed as biologically plausible\cite{honey2017switching}
and also plausible in physical systems\cite{poole2017chemical,stern2021supervised}. However, a learning rule of this kind that compares two different states raises some challenges. Experiencing the two states and updating weights in sequence, i.e., $\Delta w_{ij} = - \epsilon s_{ij}^{\rm free}$, followed by $\Delta w_{ij} = \epsilon s_{ij}^{\rm clamped}$, will require a very small learning rate $\epsilon$ to avoid convergence problems since the difference $s_{ij}^{\rm clamped} - s_{ij}^{\rm free}$ can be much smaller than either of the two terms\cite{behera2006adaptive}. Further, this approach requires globally switching the system between Hebbian and anti-Hebbian learning rules.  One natural option then is to store information about the free state synaptic currents $s_{ij}^{\rm free}$ locally at each synapse during the free state - without making any weight updates - and then store information about the clamped state currents $s_{ij}^{\rm clamped}$, again without making any weight updates, and then after a cycle of such states, update weights based on Eq. \ref{eqn:wijfreeclamped} using on the stored information.  While this approach is natural when implementing such training on a computer,
it imposes multiple requirements on the physical or biological system: (1) local memory units at each synapse that stores information about $s^{\rm free}_{ij},s^{\rm clamped}_{ij}$ before $w_{ij}$ are updated, (2) a global signal informing the system whether the current state corresponds to free or clamped state (since 
the corresponding $s_{ij}$ must be stored with signs, given the form of Eq. \ref{eqn:wijfreeclamped}). These requirements can limit the relevance of contrastive training to natural physical and biological systems, even if they can naturally update parameters through Hebbian-like rules\cite{stern2023learning,stern2020supervised,evans2022pattern,pashine2019directed,xie2003equivalence}. 

\section*{Temporal Contrastive learning using implicit memory}
We investigate an alternative model of implicit non-equilibrium memory, Temporal Contrastive Learning (TCL) in a synapse (or any equivalent physical learning degree of freedom\cite{stern2023learning}). In our model,  clamped or free states information is not stored anywhere explicitly; instead, non-equilibrium dynamics at each synapse results in changes to $w_{ij}$ that are based on the difference between past and present states.

Consider a neural network driven to experiencing a temporal sequence of inputs (e.g., smoothly interpolating between free and clamped states according to a periodic temporal protocol) which induces a synaptic current $s_{ij}(t)$ at synapse $ij$. 
For example, the Hebbian rule (`fire together, wire together') is based on a signal $s_{ij}(t) = \eta_i(t) \eta_j(t)$ where $\eta_i$ is the firing rate of neuron $i$\cite{gerstner2002mathematical, lowel1992selection}.  
The conventional Hebbian rule assumes an instantaneous update of weights with the current value of the synaptic signal,
\begin{equation}
\frac{d w_{ij}}{dt} = g(s_{ij}(t)), 
\end{equation}
where $g$ is a system-dependent non-linear function\cite{gerstner2002mathematical}.

In contrast, we consider updates based on an implicit memory of recent history $s_{ij}(t)$, 
with the memory encoded by convolution with a kernel $K(t-t')$.
\color{black}This kernel convolution arises through the underlying physical dynamics in diverse physical and biological systems to be discussed later; in this approach, there is no explicit storage of the value of the signal $s_{ij}(t)$ at each point in time in specialized memory. 
\color{black}
The memory kernel, together with the past signal, characterizes the response of each synapse to the history of signal values:
\begin{equation}
u_{ij}(t) = \int_{-\infty}^{t} K(t-t') s_{ij}(t') \,dt' 
\end{equation}
\begin{equation}
\frac{d w_{ij}}{dt} =  g(u_{ij}(t))\,,
\label{eqn:kernel}
\end{equation}
where the nonlinearity $g$ may be system-dependent. 

While most physical and biological systems have some form of memory, the most common memory is described by monotonic kernels, \color{black}for example \color{black}$K(t-t') \sim e^{-(t-t')/\tau_K}$. But a broad class of non-equilibrium systems exhibit memory with a non-monotonic kernel with both positive and negative lobes,  \color{black}for example \color{black} $K(t-t') \sim e^{- (t-t')/\tau_K} f(t-t')$, with $f(t-t')$ a polynomial\cite{tu2008modeling,tu2013quantitative,lan2012energy,becker2015optimal,tjalma2023trade,xie1999spike,riviere2023plants,celani2010bacterial,mattingly2021escherichia}. 
Such kernels arise naturally in numerous physical systems, the prototypical example being integral feedback\cite{doyle2013feedback} in analog circuits.

A key intuition for considering such an update rule can be seen in the response of the system to a step-function signal; when convolved with a non-monotonic kernel, the constant parts of the signal produce a vanishing output.
However, at the transition from one signal value to another, the convolution produces a jump with a timescale inherent to the kernel (Fig. \ref{fig:1schematic-pt2}B).
As such, the convolution in effect produces the finite-time derivative of the original signal. 

Finally, $g(x)$ in Eq. \ref{eqn:kernel} can represent any non-linearity that suppresses small inputs $x$ relative to larger $x$, e.g., 
\begin{equation}
\label{eqn:nonlinearity}
g(x) = 
 \begin{cases} 
     x  & \vert x \vert \geq \theta_g \\
      0  & \vert x \vert < \theta_g \\
   \end{cases}~~.
\end{equation}
This non-linearity allows for differentiation between fast and slow changes in synaptic signal.

\color{black}


The weight update of the system described thus far does not change in time, e.g., between clamped and free states; synapses are always updated with the same fixed local learning rule in Eq. \ref{eqn:kernel}. 

The only time-dependence comes from training examples being presented in a time-dependent way. For simplicity, we consider a sawtooth-like training protocol, where input and output neurons $\eta_i(t)$ alternate between free and clamped states over time (Fig. \ref{fig:1schematic-pt2}C).
In this sawtooth protocol, the free-to-clamped change is fast (time $\tau_f$) while the clamped-to-free relaxation is slow (time $\tau_s > \tau_f$); see Appendix \ref{app:kernel_constraints} for further detail.
\color{black}

Such a time-dependent sawtooth presentation of training examples $\eta_i(t)$ induces a sawtooth synaptic signal $s_{ij}(t)$. We can compute the resulting change in weights $\Delta w_{ij}$, which is Eq. \ref{eqn:kernel} integrated over one cycle of the sawtooth, assuming slow weight updates:
\begin{equation}\label{eqn:w}
    \Delta w_{ij}  = \epsilon~ \int_0^{\tau_f + \tau_s} g\left(\int_{-\infty}^{t} K(t-t') s_{ij}(t') \,dt' \right)\, dt ~,
\end{equation}
with $\epsilon$ a learning rate.
As shown in Appendix \ref{app:kernel_constraints}, in the specific limit of protocols with $\tau_K \ll \tau_f \ll \tau_s$:
\begin{equation}
\Delta w_{ij} \approx \epsilon (s_{ij}^{\rm clamped} - s_{ij}^{\rm free}).
\end{equation}

Intuitively, the kernel $K$ computes the approximate (finite) time derivative of $s_{ij}(t)$; \color{black} this is what requires the kernel timescale $\tau_K \ll \tau_f$. \color{black}
The rapid rise of the sawtooth from free to clamped states results in a large derivative that exceeds the threshold $\theta_g$ in $g(x)$ and provides the necessary contrastive learning update. The slow relaxation from clamped to free has a small time-derivative that is below $\theta_g$. \color{black} The ability to distinguish free-to-clamped versus clamped-to-free transitions  requires $\tau_f \ll \tau_s$ as shown in Appendix A.
\color{black}


Consequently, 
the fast-slow sawtooth protocol allows systems with non-equilibrium memory kernels to naturally learn through contrastive rules by comparing free and clamped states over time. Note that our model here does not include an explicit memory module that stores the free and clamped state configurations but rather exploits the memory implicit in local feedback dynamics at each synapse. 

\color{black}


\section*{Related Work Summary}

Several works have proposed ways in which contrastive learning can be generated in natural and engineered systems, such as: utilizing explicit memory\cite{yi2023activity,laydevant2024training}; having two globally coordinated phases each with low learning rates\cite{williams2023flexible}; having two copies of the system\cite{dillavou2022demonstration,dillavou2023machine,wycoff2022desynchronous}; having two physically distinct kinds of signals\cite{anisetti2023learning};  or using continually-running oscillations in the learning rules\cite{baldi1991,anisetti2024frequency}.
All these methods require globally switching the system between two phases or exploit hardware-specific mechanisms. The closest to the TCL approach here is `continual EP'\cite{ernoult2020equilibrium}, but it also requires globally switching between phases of learning.
See Appendix \ref{app:related_work} for a more in-depth review of these related works.

Our rule superficially resembles spike-timing dependent plasticity
(STDP)\cite{xie1999spike,bi2001synaptic,bengio2017stdp}, in particular how our TCL weight update has a representation involving a non-monotonic kernel $K$. 
While STDP-inspired rules\cite{martin2021eqspike,burkitt2004spike,moraitis2020optimality} and other rules involving competing Hebbian updates with inhibitory neurons\cite{krotov2019unsupervised,journe2022hebbian} have shown great promise as local learning paradigms, they are adapted to settings where synapses are asymmetric and can distinguish differentials between the timings of pre- and post-synaptic neuronal activations.
Our rule only involves signals at the same moment in time $t$ and can only result in symmetric interactions $w_{ij}$. 
We are not aware of any direct relationship between work on STDP rules and the proposal here; see Appendix \ref{app:related_work} for further detail.
\color{black}

\section*{Performance on MNIST}

\begin{figure}
\includegraphics[width=.95\linewidth]{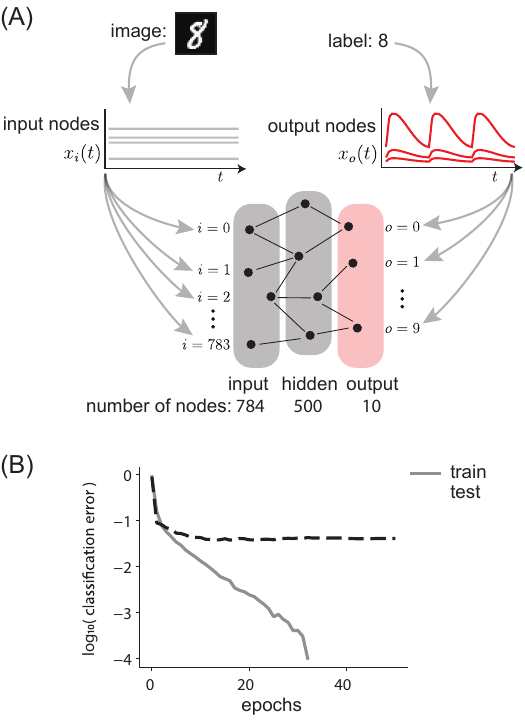}
\caption{\textbf{Temporal Contrastive Learning in a neural network for MNIST classification.} (A) We train a neural network with three types of neurons: input, hidden, and output. During training time, neurons in the input layer are set to internal states $x_i$ based on input MNIST digits (gray curves). But output neuron states $x_o$ are modulated between being nudged to the desired output (i.e., correct image label) and being free in a sawtooth-like protocol (red curves). Synaptic weights are updated using a memory kernel as in Eq. \ref{eqn:sin_kernel} with timescale $\tau_K = .1$; length of one free-clamped cycle $\tau_f + \tau_s = 1, \tau_f = .1$. See Appendix \ref{app:multi_synapse} for protocol details. 
(B) Average train classification error as a function of epochs of training (gray curve). Each epoch involves a pass through the 10000 MNIST entries in the training set, where for each MNIST entry, network weights are updated based on a single cycle of the sawtooth protocol.
Test error (dashed black curve) is computed on 2000 entries in a test set. 
}
\label{fig:mnist_example}
\end{figure}

By coupling together a network of synapses with non-equilibrium memory kernels, we were able to train a neural network capable of classifying MNIST. \color{black}In particular, we adapted the Equilibrium Propagation (EP) algorithm\cite{scellier2017equilibrium}, a contrastive learning-based method that `nudges' the system's state towards the desired state, rather than clamping it as is done in standard contrastive Hebbian learning (CHL) \citep{movellan1991contrastive}. We used EP instead of CHL due to its better properties and its superior performance in practice \citep{scellier2023energy,laborieux2022holomorphic}.  
EP makes weight updates of the form:
\begin{equation}
\Delta w_{ij} = \epsilon (s_{ij}^{\rm nudge} - s_{ij}^{\rm free}).
\end{equation}
Normally the EP method requires storing the states $s_{ij}^{\rm nudge}$ and $s_{ij}^{\rm free}$ in memory, computing the difference and then updating weights $w_{ij}$; our proposed TCL method will accomplish the above EP weight update, without explicitly storing and retrieving those states.

In order to process MNIST digits, we utilize a network architecture with three types of symmetrically-coupled nodes. 
Each node carries internal state $x$ and activation $\eta = \text{clip}(x,0,1)$. Nodes belong either to a 784-node input layer (indexed by $i$), a 500-node hidden layer (indexed by $h$), or a 10-node output layer (indexed by $o$). Nodes are connected by synapses only between adjacent layers ($i$ and $h$, $h$ and $o$), with no skip- or lateral-layer couplings.
The neural network dynamics minimize the energy:
\begin{equation}\label{eqn:nn_energy}
E(x) = \frac{1}{2}\Sigma_n x_n^2 - \frac{1}{2}\Sigma_{n, m} w_{nm} \eta_n \eta_{m} - \Sigma_n b_n \eta_n,
\end{equation}
where $b_n$ is the bias of node $n$, $w_{nm}$ is the weight of the synapse connecting nodes $n$ and $m$, and the indices $n, m$ run over the node indices of all layers $i,h,o$. 

We represent each $784$-pixel grayscale MNIST image as a vector $v^{\text{image}} = \{v^{\text{image}}_i\}_{i = 0, \ldots, 783}$. For each MNIST digit $v^{\text{image}}$, we hold the states of the 784 input nodes at constant values over time, $x_i(t) = v^{\text{image}}_i$. 
For inference, we allow the hidden- and output-layer activations to adjust in response to the fixed input nodes, minimizing Eq. \ref{eqn:nn_energy}.
Once a steady-state is reached, the network prediction is given by looking at the states of the 10 output nodes, $\{x_o\}_{o = 0, \ldots, 9}$, and interpreting the index of the maximally activated output node as the input image label.

During inference, the network minimizes an energy function which does not vary in time, subject to the constraint that $x_i(t) = v^{\text{image}}_i$. During training, the same constraint $x_i(t) = v^{\text{image}}_i$ applies, but our network instead is subjected to a time-varying energy function:
\begin{equation}\label{eqn:nn_energy_augment}
F(x;t) = E(x) + \frac{\beta(t)}{2} \Sigma_{o=0}^{9}(x_{o} - v^{\text{label}}_o)^2 .
\end{equation}
Here, $v^{\text{label}}$ is the one-hot encoding vector for the corresponding label of the MNIST digit.
The time-dependent training protocol $\beta(t)$ is the asymmetric sawtooth function which smoothly interpolates between $0$ and a maximal value $\beta_{max} < 1$. One portion of the sawtooth is characterized by the fast timescale $\tau_f$, and the other portion by the slow timescale $\tau_s$. We assume that both these timescales are quasi-static compared to any system-internal energy relaxation timescales\cite{albash2018adiabatic}.
This restriction to adiabatic protocols is a potential limitation to contrastive methods in general; details depend on the system and some works present viable workarounds\cite{hinton2002training,stern2022physical}.

Each time the training protocol $\beta(t)$ completes a cycle, the network weights are updated according to Eq. \ref{eqn:w}, with $\eta_n \eta_m$ as the necessary synaptic current $s_{nm}$.
After completing multiple sawtooth cycles, we switch the inputs $x_i$ to a new MNIST image and repeat the training process of manipulating the $x_o$ through the time-varying energy function $F(x;t)$. See Appendix \ref{app:multi_synapse} for further detail, including specifications of $K$ and $g$ for Eq. \ref{eqn:w}.


\color{black}

We find that, after training for 35 epochs, our classification error drops to 0, and we achieve an accuracy of 95\% on our holdout test dataset (Fig. \ref{fig:mnist_example}B).
Our results demonstrate the feasibility of performing contrastive learning in neural networks without requiring explicit memory storage, but leave open the question of limitations on our approach.
We consider performance limitations at the level of a single synapse in the following sections.

\section*{Speed-accuracy tradeoff}

\begin{figure}
\includegraphics[width=\linewidth]{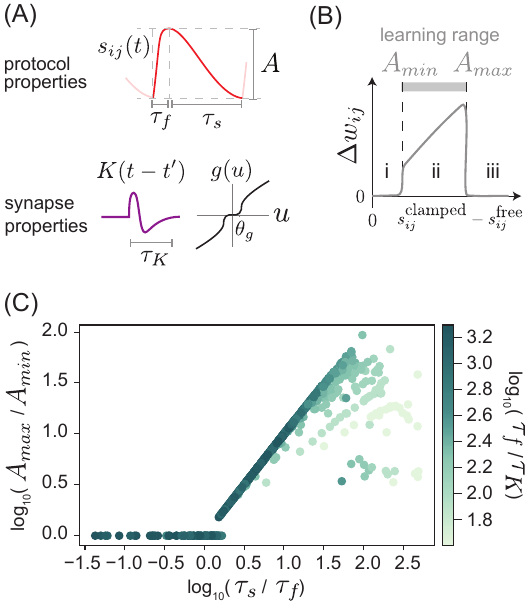}
\caption{\textbf{Tradeoff between time and dynamic range of synaptic currents over which contrastive learning is approximated.} (A) The sawtooth training protocol $s_{ij}(t)$ with amplitude $A$ approximates contrastive learning in specific regimes of protocol timescales ($\tau_s,\tau_f$) relative to nonlinearity threshold $\theta_g$ and synaptic memory kernel timescale $\tau_{K}$. (B) The weight update $\Delta w_{ij}$ is proportional to the signal amplitude $s^{\rm clamped} - s^{\rm free}$ as desired within a range of amplitudes (regime ii) for given timescales, shown here for $\tau_{f}$ = 2, $\tau_{s}$ = 10, $\tau_{K}$ = 1/20, $\theta_{g}$ = 3. (C) 
Learning range (defined in (B) as the dynamic range $A_{max}/A_{min}$ for which the sawtooth protocol approximates contrastive Hebbian learning) plotted against $\tau_s/\tau_f$.
Signal timescales ($\tau_s,\tau_f$) place a limit on maximal learning range which increases with increasing $\tau_s/\tau_f$. Protocols with higher $\tau_{f}/\tau_{K}$ approach this limit. 
}
\label{fig:speed_limitations}
\end{figure}

The learning model proposed here relies on breaking the symmetry between free and clamped using timescales in a sawtooth protocol (Fig. \ref{fig:1schematic-pt2}C). The expectation is that our proposal for approximating the difference between free and clamped states works best when this symmetry-breaking is large; i.e. in the limit of slow protocols. 

We systematically investigated such speed-accuracy trade-offs inherent to the temporal strategy proposed here. First, we fix synaptic parameters $\theta_g,\tau_K$ and protocol parameters $(\tau_s,\tau_f)$ and consider protocols of varying amplitude $A = s^{\rm clamped} - s^{\rm free}$ (Fig. \ref{fig:speed_limitations}A). 
\color{black} The protocol we use can be written in the form:
\begin{equation}\label{eqn:saw}
s_{ij}(t) = 
 \begin{cases} 
     \frac{A t}{\tau_f} + \overline{s_{ij}} - \frac{A}{2} & t\leq \tau_f \\
      \overline{s_{ij}} + \frac{A}{2} - \frac{A  (t-\tau_f)}{\tau_s} & \tau_f < t\leq \tau_s \\
   \end{cases}
\ ~,
\end{equation}
with $\overline{s_{ij}}$ the average of the clamped and free states.
For concreteness, we fix the kernel $K$ to be of the form $K(t-t') \sim e^{- (t-t')/\tau_K} f(t-t')$, with the exact expression for the polynomial $f$ given in Appendix \ref{app:kernel_coefficients}.
\color{black}
We compute $\Delta w_{ij}$ for these protocols and plot against amplitude $A$ in Fig. \ref{fig:speed_limitations}B. A line with zero intercept and slope $1$ indicates a perfect contrastive Hebbian update. We see that the contrastive Hebbian update is approximated for a regime of amplitudes $A_{min} < A < A_{max}$, where $A_{min},A_{max}$ are set by the requirement that $\theta_g$ separates the rate of change in the fast and slow sections of the protocol (Fig. \ref{fig:speed_limitations}B). 

We determined this dynamic range $A_{max}/A_{min}$ for protocols of different timescales $\tau_s,\tau_f$, keeping the kernel fixed but always choosing 
a threshold $\theta_g$ which provides an optimal dynamic range for the synapse output, given a maximal amplitude magnitude (see Appendix \ref{app:kernel_constraints}).
Choosing $\theta_g$ this way, we find the following trade-off equation,
\begin{equation}
    \tau_s > \tau_f \frac{A_{max}}{A_{min}},
\end{equation}
that is, the slower the speed of the down ramp, the larger the range of amplitudes over which contrastive learning is approximated (Fig. \ref{fig:speed_limitations}C).

Finally, we also need $\tau_K \ll \tau_f$; intuitively, the kernel $K(t-t')$ is a derivative operator but on a finite timescale $\tau_K$; thus the fast ramp must last longer than this timescale so that $u_{ij}(t)$ can reflect the derivative of $s_{ij}(t)$. 
As $\tau_f \to \tau_K$, the contrastive Hebbian approximation breaks down (see Appendix \ref{app:kernel_constraints}).
By setting $\tau_f/\tau_K$ to be just large enough to capture the derivative of $s_{ij}(t)$, we find that the central trade-off to be made is between having a longer $\tau_s$ and a larger dynamic range $A_{max}/A_{min}$ (Fig. \ref{fig:speed_limitations}C). 

\section*{Energy dissipation cost of non-equilibrium learning}

Our mechanism is fundamentally predicated on memory as encoded by the non-monotonic kernel $K(t-t')$. At a fundamental level, such memory can be linked to non-equilibrium dynamics to provide a Landauer-like principle for learning\cite{bennett2003notes,lan2012energy}.
Briefly, we will show here that learning accuracy is reduced if the kernel has non-zero area $I = \int_{0}^{\infty} K(t)dt$; we then show, based on prior work\cite{tu2008nonequilibrium,lan2012energy}, that reducing the area of kernels to zero requires increasingly large amounts of \color{black}energy dissipation.
In other words, to perform increasingly accurate inference, systems need to consume more energy (e.g. electrical energy in neuromorphic systems, ATP or other chemical fuel in molecular systems), which is then dissipated as heat.
\color{black}

\begin{figure}
\includegraphics[width=\linewidth]{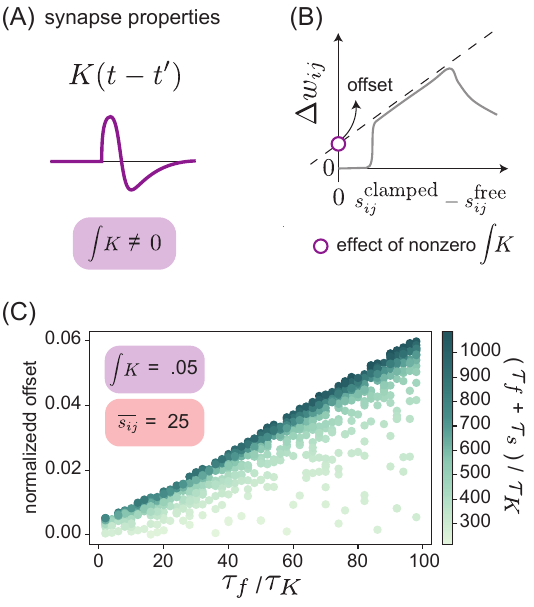}
\caption{\textbf{Kernels with nonzero area limit contrastive learning.} (A) Memory kernels with nonzero area ($I = \int_{0}^\infty K dt$) are sensitive to the time-average synaptic current $\overline{s_{ij}}$. (B) For protocols with kernels with nonzero area, $\Delta w_{ij}$ is no longer proportional to $s^{\rm clamped} - s^{\rm free}$ within the range of performance but instead has a constant offset. Offset is the y-intercept of the line of best fit for the linear portion of the weight update curve. (C) Offset from contrastive learning (defined in (B)) for protocols of varying $\tau_f,\tau_s$, evaluated for a fixed memory kernel of timescale $\tau_K$ and non-zero area.  Offset is positively correlated with $\tau_{f}$, with overall longer protocols (high $\tau_{f} + \tau_{s}$) offset the most. Here the average signal value $\overline{s_{ij}}$ = 25; offset is normalized by $\overline{s_{ij}}$.
}
\label{fig:nonzero_area_kernels}
\end{figure}

We first consider the problem at the phenomenological level of kernels $K$ with non-zero integrated area $I = \int_{0}^{\infty} K(t) dt$ (Fig. \ref{fig:nonzero_area_kernels}A, Appendix \ref{app:AUC_kernels}).
In the limit of small $I$ and $\tau_K$, and large $\tau_s$, the weight update $\Delta w$ to a synapse experiencing a current $s_{ij}(t)$ reduces to:
\begin{equation}
    \epsilon^{-1} \Delta w_{ij} = (s_{ij}^{\rm clamped} - s_{ij}^{\rm free}) +  \frac{\tau_fI}{2}(s_{ij}^{\rm free} + s_{ij}^{\rm clamped})
\end{equation}
where $\tau_f$ is the fast timescale of the protocol. See Appendix \ref{app:AUC_kernels} for further detail. Hence, the response of the synapse with a non-zero area deviates from the ideal contrastive update rule by the addition of an offset proportional to the average signal value (Fig. \ref{fig:nonzero_area_kernels}B).

Indeed, when we fix integrated kernel area and mean signal value, we find that increasing protocol lengths (both $\tau_f$ and $\tau_s$, due to finite $\tau_K$) result in a larger offset, and hence further deviations from the ideal contrastive update (Fig. \ref{fig:nonzero_area_kernels}C).
This constraint is in contrast to the metric used for assessing performance of kernels with zero integrated area, where we found that longer protocols enabled learning over a wider dynamic range of amplitudes. This result suggests that optimal protocols will be set through a balance between the desire to minimize the offset created by non-zero kernel area, and the desire to maximize the dynamic range of feasible contrastive learning.

We now relate the kernel area to learning accuracy to provide a Landauer-like\cite{bennett2003notes} relationship between energy dissipation and contrastive learning. To understand this fundamental energy requirement, we 
consider a microscopic model of reversible non-equilibrium 
feedback. Such models have been previously studied to understand the fundamental energy cost of molecular signal processing during bacterial chemotaxis\cite{lan2012energy,murugan2017topologically}. In brief, non-monotonic memory kernels $K$ require breaking detailed balance and thus dissipation; further, reducing the area $I$ to zero requires increasingly large amounts of  dissipation. 

As shown in Refs. \cite{lan2012energy,murugan2017topologically}, the simplest way to model a non-monotonic kernel with a fully reversible non-equilibrium statistical model is to use a Markov chain shaped like a `ladder network' (Appendix \ref{app:markov_chain}). The dynamics of the Markov chain are governed by the master equation:
\begin{equation}
    \frac{d}{dt} p_a = \sum_b r_{ba} p_b - p_a\sum_b r_{ab} ~.
\end{equation}
Here $r_{ab}$ are the rate constants for transitions from state $a$ to $b$, and $p_a$ is the occupancy of state $a$. Here we consider a simple Markov chain network arranged as a grid with two rows (see Fig. \ref{fig:markov_cartoon}A for a schematic, and Appendix \ref{app:markov_chain} for the fully specified network). The dynamics of the network are mainly controlled by two parameters: $s_{ij}$ and $\gamma$. As in Ref. \cite{murugan2017topologically}, the rates ($r_{up}^i, r_{down}^i$) on the vertical (red) transitions are driven by synaptic current $s_{ij}$. These rates vary based on horizontal position such that circulation along vertical edges is an increasing function of $s_{ij}$; see Appendix \ref{app:markov_chain} for the functional form of this coupling. Quick changes in $s_{ij}$ shift the occupancy $\vec{p}$, which then settles back into a steady state. Thus through $s_{ij}$ the network is perturbed by an external signal. On the other hand, the parameter $\gamma$ controls the intrinsic circulation within the network by controlling the ratio of clockwise ($r_{cw}$) to counterclockwise $(r_{ccw})$ horizontal transition rates:

\begin{figure}
\includegraphics[width=\linewidth]{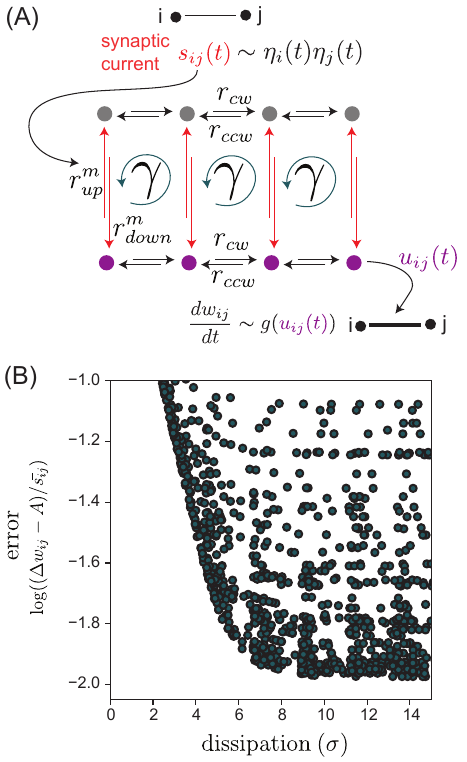}
\caption{\textbf{Dissipation cost of contrastive learning.} (A) A thermodynamically reversible Markov state model of physical systems with non-monotonic kernel $K$. Nodes represent states connected by transitions with rate constants as shown; the ratio of rate constants $r^{m}_{up}$ and $r^m_{down}$ is set by synaptic current $s_{ij}(t)$ (the input). The output $u_{ij}$ is defined as the resulting total occupancy of purple states; $u_{ij}$ is used to update weight $w_{ij}$. $g$ is a thresholding nonlinearity as in Fig. \ref{fig:speed_limitations}A. The ratio of products of clockwise to counterclockwise rate constants, $\gamma$, quantifies detailed balance breaking and determines energy dissipation $\sigma$. 
(B) Energy dissipation $\sigma$ versus error in contrastive learning rule for different choices of rate constants in (A). Error of contrastive learning is quantified by the normalized offset from ideal contrastive weight update; $\Delta w_{ij}$ is the weight update $\int_0^t\,g(u(t'))dt'$, $A$ is the signal amplitude $s_{ij}^{\rm clamped} - s_{ij}^{\rm free}$,  and $\bar{s_{ij}}$ is the average signal value $({s_{ij}^{\rm clamped} + s_{ij}^{\rm free}})/2$ for a sawtooth signal (see Fig. \ref{fig:nonzero_area_kernels}B). 
Bounded region indicates greater dissipation is required for lower offset and better contrastive learning performance.
}
\label{fig:markov_cartoon}
\end{figure}

\begin{equation}
    \gamma = \frac{r_{cw}}{r_{ccw}}.
\end{equation}
Then $u_{ij}$ is taken to be the occupancies $\sum_i p_i$ over all nodes $i$ along one rail of the ladder (Fig. \ref{fig:markov_chain}A). 
The parameter $\gamma$ is particularly key in quantifying the breaking of detailed balance. When $\gamma \rightarrow 0$, i.e., when this network is driven out of equilibrium, the microscopic dynamics of this Markov chain model provide a memory kernel $K$ suited for contrastive learning. In particular, the response of the Markov chain to a small step function perturbation in $s_{ij}$ (i.e., to vertical rate constants) results in $u_{ij}(t) = \int_{-\infty}^{t} K(t-t') s_{ij}(t')dt'$ with a memory kernel $K(t-t')$ much as needed for Fig. \ref{fig:1schematic-pt2}B. The form of $K$ can then be extracted from $u_{ij}$ by taking its derivative and normalizing.

\color{black}In the Markov chain context, we can rigorously compute energy dissipation $\sigma = \sum_{i > j}({r}_{ij}p_j - {r}_{ji}p_i)\ln(\frac{{r}_{ij}p_j}{{r}_{ji}p_i})$. 
\color{black}
For the particular ladder network studied here, dissipation takes the following form (in units of $kT$):
\begin{equation}\label{diss}
\sigma = \sigma_v + \sum_{i = 0}^{8}(p_{i} - \gamma p_{i+1})\ln(\frac{p_{i}}{\gamma p_{i+1}})
\end{equation}
Where $\sigma_v$ is dissipation along vertical connections, and the node occupations $p_i$ (excluding corners, which are accounted for by $\sigma_v$) are indexed in counterclockwise order. The dissipation is a decreasing function of $\gamma \in [0,1]$. This dissipation (at steady state), is zero if $\gamma = 1$ (i.e., detailed balance is preserved) and non-zero otherwise. Markov chains with no dissipation produce monotonic memory kernels which do not capture the finite time derivative of the signal. 

However, for a finite amount of non-equilibrium drive $\gamma$, the generated kernel will generally have both a positive and negative lobe and have lower but non-zero integrated area $I$, leading to imperfect implementation of the contrastive learning rule, offset as a result of non-zero area. 
Larger dissipation of the underlying Markov chain, e.g., by decreasing $\gamma \rightarrow 0$, will lower the learning offset and cause the kernel convolution to approach an exact finite time derivative. Fig. \ref{fig:markov_chain}B shows the increased dissipation required to reach lower offset.

In summary, Landauer's principle\cite{bennett2003notes} relates the fundamental energy cost of computation to the erasure of information (`forgetting') inherent in most computations. Our work here provides a similar rational for why a physical system learning from the environment must similarly dissipate energy - contrastive learning fundamentally requires comparing states seen across time. Hence such learning necessarily requires temporarily storing information about those states and erasing that information upon making weight updates. While the actual energy dissipation can vary depending on implementation details in a real system, our kernel-based memory model allows for calculating this dissipation cost associated with learning in a reversible statistical physics model.

\section*{Realizing memory kernels through integral feedback}

\begin{figure*}
\includegraphics[width=0.95\linewidth]{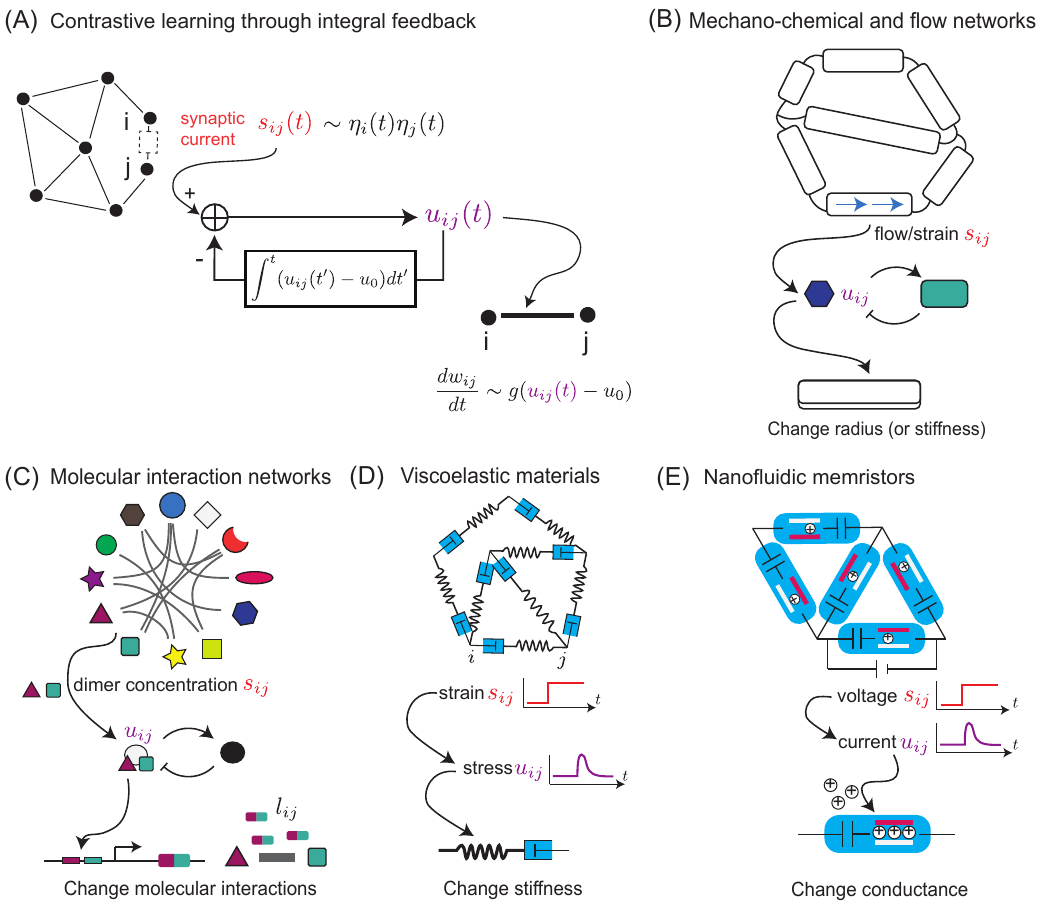}
\caption{
\textbf{Integral feedback allows diverse Hebbian-capable systems to achieve contrastive learning.}
(A) Synaptic current $s_{ij}$ in synapse $ij$ impacts a variable $u_{ij}(t)$ which is under integral feedback control. That is, the deviation of $u_{ij}(t)$ from a setpoint $u_0$ is integrated over time and fed back to $u_{ij}$, causing $u_{ij}(t)$ to return to $u_0$ after transient perturbations due changes in $s_{ij}(t)$. We can achieve contrastive learning by updating synaptic weights $w_{ij}$ using $u_{ij}(t)$, as opposed to Hebbian learning by updating $w_{ij}$ based on $s_{ij}(t)$. 
(B) Mechanical or vascular networks can undergo Hebbian learning if flow or strain produce molecular species (blue hexagon $u_{ij}$) that drive downstream processes that modify radii or stiffnesses of network edges. But if those blue molecules are negatively autoregulated (through the green species as shown), these networks can achieve contrastive learning. 
(C) Molecular interaction networks can undergo Hebbian learning if the concentrations of dimers $s_{ij}$, formed through mass-action kinetics, drives expression of linker molecules $l_{ij}$ (here, purple-green rectangles) that mediate binding interactions between monomers $i,j$. But if transcriptionally active dimers $u_{ij}$ additionally stimulate regulatory molecules (black circle) which inhibit $u_{ij}$, then levels of linker molecules $l_{ij}$  will provide interactions learned through contrastive rules.
(D) Stresses in networks of viscoelastic mechanical elements (dashpots connected in series to springs) reflect the time-derivatives of strains due to relaxation in dashpots. These stresses can be used to generate contrastive updates of spring stiffnesses.
(E) Nanofluidic memristor networks can undergo Hebbian updates by changing memristor conductance in response to current flow. However, if capacitors are added as shown, voltage changes $s_{ij}$ across a memristor results in a transient current $u_{ij}$;  conductance changes due to these currents $u_{ij}$ result in contrastive learning.}
\label{fig:implementation}
\end{figure*}

We have established that contrastive learning can arise naturally as a consequence of non-equilibrium memory as captured by a non-monotonic memory kernel $K$. Here, we argue that the needed memory kernels $K$ in turn arise naturally as a consequence of integral feedback control in a wide class of physical systems. 
\color{black}Hence many simple physical and biological systems can be easily modified and manipulated to undergo contrastive learning.
\color{black}

Integral feedback control\cite{yi2000robust,aoki2019universal,nemenman2012information,shimizu2010modular} is a broad homeostatic mechanism that adjusts the output of a system based on measuring the integral of error (i.e., deviation from a fixed point). For example, consider a variable $u$ that must be held fixed at a set point $u_0$ despite perturbations from a signal $s$. Integral feedback achieves such control by up- or down-regulating $u$ based on the integrated error signal $u-u_0$ over time,
\begin{eqnarray}\label{eqn:u}
    \tau_u \frac{du}{dt}  &=&  - u + k(s(t) - m) \\\label{eqn:m}
    \tau_m \frac{dm}{dt} & = & (u - u_{0})
\end{eqnarray}


Here, we rely on a well-known failing of this mechanism\cite{yi2000robust,lan2012energy} - if $s(t)$ changes rapidly (e.g., a step function), the integral feedback will take a time $\tau_{K}$ to restore homeostasis; in this time, $u(t)$ will rise transiently and then fall back to $u_0$ as shown in Fig. \ref{fig:1schematic-pt2}B. 
The general solution of Eqs. \ref{eqn:u}, \ref{eqn:m} is of the form of a driven damped oscillator driven by the time derivative of $s(t)$ \cite{yi2000robust,doyle2013feedback}. 
The kernel $K(t)$ can be written as the solution for forcing $s(t) = \delta(t)$.
In the overdamped regime $k\tau_u \ll \tau_m$, the kernel is of the form $K(t) = e^{-bt}\left(\text{cosh}(\omega t)-\frac{b}{\omega}\text{sinh}(\omega t) \right)\Theta(t)$, 
where $ b = \frac{1}{2\tau_u}, \, \omega = \sqrt{\frac{k}{\tau_u\tau_m} - \frac{1}{4\tau_u^2}}$, $\Theta$ is a unit step function, and we have chosen $u_0 = 0$. This form of $K(t)$ is a non-monotonic function with both a positive and negative lobe, which returns to the set point $u_0 = 0$ after a perturbation. Hence simple integral feedback can create memory kernels of the type required by this work. We propose a large class of physical systems capable of Hebbian learning can be promoted to contrastive learning by a layer of integral feedback as shown Fig. \ref{fig:implementation}A.

\textbf{Learning in mechano-chemical systems.}
Numerous examples of biological networks release chemical signals in response to mechanical deformations $s_{ij}$; e.g., due to shear flow forces in \textit{Physarum polycephalum}\cite{marbach2023vein,kramar2021encoding,alim2017mechanism} or strain in cytoskeletal networks\cite{cavanaugh2020rhoa,cavanaugh2020adaptive}. If the chemical signals then locally modify the elastic moduli or conductances of the network, these systems can be viewed as naturally undergoing Hebbian learning. 

Our proposal suggests that these systems can perform contrastive learning  if the chemical signal is integral feedback regulated (Fig. \ref{fig:implementation}B). For example, if the molecule released by the mechanical deformation is under negative autoregulation, then that molecule's concentration $u_{ij}(t)$ is effectively the time derivative of the mechanical forcing of the network, $\dot{s}_{ij}(t)$. \color{black}A simple model for negative autoregulation is:
\begin{eqnarray}\label{eqn:u_7b}
    \tau_u \frac{du}{dt}  &=&  - u + k_as(t) - k_im\frac{u}{u_s + u} \\\label{eqn:m_7b}
    \tau_m \frac{dm}{dt} & = & u - u_0~~,
\end{eqnarray}
Here, $u$ is the level of activated molecules (e.g., phosphorylated form); we assume that phosphorylation is driven with rate $k_a$ by the strain molecule $s$. However, excess levels of $u$ relative to a baseline $u_0$ leads to build up of another molecular form $m$ (e.g., methylated molecules in the case of chemotaxis\cite{doyle2013feedback,barkai1997robustness}). If we assume that $m$ then deactivates (or dephosphorylates as in chemotaxis) $u$ with rate $k_i$, we obtain an control loop as long as $u$ remains above a small saturating threshold $u_s$. 
These dynamics provide integral feedback control over $u$\cite{qian2018realizing}, and allow for computing the time-derivative of strain $s$.
\color{black}

Therefore, altering the radius in flow networks or stiffness in elastic networks based on $u_{ij}(t)$ will now lead to contrastive learning. The same mechano-chemical principles can also be exploited in engineered systems such as DNA-coupled hydrogels\cite{cangialosi2017dna}; in these systems, DNA-based molecular circuits\cite{scalise2019controlling, schaffter2019building, fern2017dna, fern2017design} can implement the needed integral feedback while interfacing with mechanical properties of the hydrogel. Such chemical feedback could be valuable to incorporate into a range of existing metamaterials with information processing behaviors\cite{kwakernaak2023counting,arinze2023learning}. 


\textbf{Learning in molecular systems.}
Molecular systems can learn in a Hebbian way through `stay together, glue together' rules, analogous to the `fire together, wire together' maxim for associative memory in neural networks\cite{murugan2015multifarious,zhong2017associative,evans2022pattern,poole2017chemical,poole2023autonomous,owen2023design}. In the example shown in Fig. \ref{fig:implementation}C, temporal correlations between molecular species $i$ and $j$ would result in an increased concentration of `linker' molecules $l_{ij}$ that will enhances the effective interaction between $i$ and $j$. This mechanism exploits dimeric transcription factors\cite{zhu2022synthetic,parres2023principles}; an alternative proposal\cite{evans2022pattern} involves proximity-based ligation\cite{moerman2022simple} as used in DNA microscopy\cite{weinstein2019dna}. 

We can promote these Hebbian-capable systems to contrastive learning-capable systems by including a negative feedback loop. In this scheme, monomers $i$,$j$ form dimers $s_{ij} \propto c_i c_j $ dictated by mass-action kinetics, as in Hebbian molecular learning\cite{murugan2015multifarious,evans2022pattern}. We now assume that these dimers can form a transcriptionally active component $u_{ij}$ by binding an activating signal (e.g. shown as a circle in Fig.\ref{fig:implementation}C). 

Here, molecular learning\cite{murugan2015multifarious,evans2022pattern} takes place as follows:  monomers $i$,$j$ form a compound dimer $u_{ij}$ whose concentration is dictated by mass-action kinetics $s_{ij}\propto c_i c_j$; these $u_{ij}$ drive the transcription of a linker molecule $l_{ij}$ that mediates Hebbian-learned interactions between $i$ and $j$.
To achieve Temporal Contrastive Learning, we now additionally allow activated dimers $u_{ij}$ to produce a regulating signal $m_{ij}$ which deactivates or degrades $u_{ij}$ during training time\cite{alon2019introduction,zhu2022synthetic}. At test time, the resulting  interaction network for monomers $i$,$j$, mediated by linker levels $l_{ij}$, will reflect interactions learned through contrastive rules.
\color{black}
This scheme can be quantitatively written as:
\begin{eqnarray}\label{eqn:u_7c}
    \tau_u \frac{du}{dt}  &=&  - u + k_a s(t) - k_i m \frac{u}{u_s + u}  \\\label{eqn:m_7c}
    \tau_m \frac{dm}{dt} & = & u - u_0\\\label{eqn:l_7c}
    \tau_l \frac{dl}{dt} & = & g(u-u_0)~~,
\end{eqnarray}
where $k_a$ and $k_i$ are rates of production and degradation, $u_s$ is a small saturating threshold for degredation, $u_0$ is a baseline for the production of $m$, and $g$ is a transcription-related nonlinearity for the production of linkers $l$.
\color{black}

\textbf{Learning in viscoelastic materials.}
Mechanical materials with some degree of plasticity have frequently been exploited in demonstrating how memory formation and Hebbian learning can arise in simple settings, e.g. with foams\cite{pashine2019directed,hexner2020effect}, glues\cite{arinze2023learning}, and gels\cite{scheff2021actin}.
In this setting, stress slowly softens the moduli of highly strained bonds, thereby lowering the energy of desired material configurations in response to a strain signal $s_{ij}(t)$.

We add a simple twist to this Hebbian framework in order to make such systems capable of contrastive learning; we add a viscoelastic element with a faster timescale of relaxation compared to the timescale of bond softening. In reduced-order modeling of viscoelastic materials, a common motif is that of a viscous dashpot connected to an elastic spring in series, also known as a Maxwell material unit. \color{black}Such units obey the following simple equation:
\begin{equation}\label{eqn:u_7d}
\frac{\dot{u}}{k} + \frac{u}{\gamma} = \dot{s}
\end{equation}
where $u$ is the stress in the unit, $s$ is the strain in the unit, $k$ is the Hookean modulus of the spring and $\gamma$ is the Newtonian viscosity.
In this simple model, a Maxwell unit experiences integral feedback; sharp jumps in the strain $s_{ij}(t)$ lead to stresses $u_{ij}(t)$ across the edge which initially also jump, but then relax as the viscous dashpot relieves the stress.
As such, the stress in each edge of the network naturally computes an approximate measurement of the time derivative of strain the edge experiences, $\dot{s}_{ij}(t)$.
We can see this cleanly by solving for the case where $s(t) = \dot{s}t$ with $\dot{s}$ constant, in which case $u(t) = \gamma\dot{s}(1 - e^{-\frac{k}{\gamma}t})$, in which case $u \propto \dot{s}$ after a relaxation time $\frac{\gamma}{k}$.
\color{black}

We consider a network of such units (Fig. \ref{fig:implementation}E), experiencing strains $s_{ij}(t)$ across each edge.
If the springs in this network naturally adapt their moduli $k_{ij}$ on a slow timescale in response to the stresses they experience, they would naturally implement Hebbian learning rules. With the introduction of viscoelasticity through the dashpots, the system is now capable of contrastive updates.
\color{black} Note that, in contrast to the solution from idealized Eqs. \ref{eqn:u}, \ref{eqn:m}, the material moduli $k_{ij}$ are involved both as learning degrees of freedom and as elements of the integral feedback control.\color{black}

\textbf{Learning in nanofluidic systems.}
An emerging platform for neuromorphic computing involves nanofluidic memristor networks (Fig. \ref{fig:implementation}B) driven by voltage sources\cite{robin2021modeling,robin2023long,xiong2023neuromorphic,kamsma2023iontronic}. 
Each memristic element naturally displays Hebbian behavior; as voltage drops across the memristor, ions are recruited which modify the conductance of the element.
In this Hebbian picture, the voltage drop across a network element plays the role of the synaptic current $s_{ij}$. 

Here, we make a simple modification to allow for contrastive learning, where each element in the network is composed not of a single memristor but of a capacitor and memristor connected in series.
If voltage drops are established across a compound capacitor memristor element, then voltage initially drives a current across the memristor.
However, this initial spike in current is suppressed on the timescale of the charging capacitor.
\color{black}To see this, note that as in the viscoelastic system (Eq. \ref{eqn:u_7d}), each unit has dynamics of the form:
\begin{equation}\label{eqn:u_7e}
R\dot{u} + \frac{u}{C} = \dot{s} , 
\end{equation}
where here $u$ is the current in the unit, $s$ is the voltage across the unit, $R$ is the resistance and $C$ is the capacitance.
For the case where $\dot{s}$ is constant,  $u(t) = C\dot{s}(1 - e^{-\frac{t}{RC}})$, in which case $u(t) \propto \dot{s}$ after a relaxation time $RC$.
\color{black}
Therefore, due to the feedback control from the capacitor, the voltage that the memristor develops  is $u \approx \dot{s}$.
This simple modification therefore naturally allows for nanofluidic systems which update network conductances contrastively.

\section*{Discussion}


Backpropagation is a powerful way of training neural networks using GPUs. Training based on local rules -- where synaptic connections are updated based on states of neighboring neurons -- offer the possibility of distributed training in physical and biological systems through naturally occurring processes. However, one powerful local learning framework, contrastive learning, seemingly requires several complexities; naively, contrastive learning requires a memory of `free' and `clamped' states seen over time and/or requires the learning system to be globally switched between Hebbian and anti-Hebbian learning modes over time. Here, we showed that such complexities are alleviated by exploiting non-equilibrium memory \emph{implicit} in the integral feedback update dynamics at each synapse that is found in many physical and biological systems. 

Our Temporal Contrastive Learning approach offers several conceptual and practical advantages. In comparison to backpropagation, it provides a learning algorithm in hardware where no central processor is available.
In comparison to other local learning algorithms, which may still require some digital components, our approach offers several advantages. To start, TCL does more with less. A single analog operation --- Hebbian weight update based on an integral feedback-controlled synaptic current --- effectively stores memory of free and clamped states, retrieves that information, computes the difference, and updates synaptic weights. No explicit memory element is needed. Further, since integral feedback occurs naturally in a range of systems\cite{alon2019introduction,doyle2013feedback} and we exploit a failure mode inherent to integral feedback, our approach can be seen as an example of the non-modular `hardware is the software' philosophy\cite{laydevant2023hardware} that provides more robust and compact solutions.

Our framework has several limitations quantified by speed-accuracy and speed-energy tradeoffs derived here, \color{black} 
in addition to the general concerns about adiabatic protocols in contrastive methods\cite{hinton2002training,albash2018adiabatic}.  The time costs in our scheme (e.g. those shown in Fig. \ref{fig:speed_limitations}) are inherently larger compared to other Equilibrium Propagation-like methods which use explicit memory and global signals to switch between wake and sleep, thereby avoiding slow ramping protocols. \color{black}Slower training protocols in systems with higher energy dissipation lead to better approximations of true gradient descent on the loss function. However, we note that \emph{not} performing true gradient descent is known to provide inductive biases with generalization benefits in other contexts\cite{hardt2016train,feng2021inverse}; we leave such an exploration to future work.

Darwinian evolution is the most powerful framework we know that drives matter to acquire function by experiencing examples of such function over its history\cite{watson2016can}; however, Darwinian evolution requires self-replication. Recent years have explored `physical learning'\cite{stern2023learning} as an alternative (albeit less powerful) way for matter to acquire functionality without self-replication. For example, in one molecular version\cite{murugan2015multifarious,zhong2017associative}, recently realized at the nanoscale\cite{evans2022pattern}, molecules with Hebbian-learned interactions can perform complex pattern recognition on concentrations of 100s of molecular species, deploying different molecular self-assembly on the nanoscale. Similar Hebbian-like rules have allowed physical training of mechanical systems\cite{pashine2019directed,arinze2023learning,falk2023learning} for specific functionality. This current work shows that natural physical systems can exploit more powerful learning frameworks that require both Hebbian and anti-Hebbian training\cite{stern2020supervised,stern2021supervised} by exploiting integral feedback control. Since integral feedback is relatively ubiquitous and achieved with relatively simple processes, 
the work raises the possibility that sophisticated non-Darwinian `learning' processes could be hiding in plain sight in biological systems. In this spirit, it is key to point out that other ideas for contrastive learning have been proposed recently\cite{anisetti2024frequency,anisetti2023learning}. Further, the contrastive framework is only one potential approach to learning through local rules plausible in physical systems; the feasibility of other local frameworks\cite{journe2022hebbian,hinton2022forward,lopez2023self} in physical systems remain to be explored. 

By introducing a reversible statistical physics model for the memory needed by contrastive learning, we provided a fundamental statistical physics perspective on the dissipation cost of learning. 
While specific implementations will dissipate due to system-specific inefficiencies, our work points at an unavoidable reason for dissipation in the spirit of Landauer\cite{bennett2003notes}.
Our analysis of dissipation here only focuses on the contrastive aspect of contrastive learning. Developing reversible models for other aspects of the learning process, as has been done for inference and computation extensively\cite{bennett2003notes,ouldridge2017thermodynamics,lang2014thermodynamics,lan2012energy,murugan2014discriminatory,Ehrich2023-bd,Still2012-as,bryant2023physical,mcgrath2017biochemical,ten2016fundamental}, can shed light on fundamental dissipation requirements for learning in both natural and engineered realms.

\section*{Data availability}
The data generated in this study may be found in the following figshare database: https://doi.org/10.6084/m9.figshare.25057793.

\section*{Code availibility}
Code for training neural networks to perform MNIST recognition can be found at https://github.com/falkma/ContrastiveMemory-Exp. Code for understanding tradeoffs for protocol time and dissipation can be found at https://github.com/atstrupp/ContrastiveMemory-SynapseAnalysis.

\bibliography{contrastive_learning,biblio}

\section*{Acknowledgments}
The authors thank Lauren Altman, Marjolein Dijkstra, Sam Dillavou,  Douglas Durian, Andrea Liu, Marc Miskin, Krishna Shrinivas, Menachem Stern, and Erik Winfree for discussion.
AM acknowledges support from the NSF through DMR-2239801 and by NIGMS of the NIH under award number R35GM151211. MJF is supported by the Eric and Wendy Schmidt AI in Science Postdoctoral Fellowship, a Schmidt Sciences program. This work was supported by the NSF Center for Living Systems (NSF grant no. 2317138) and by the University of Chicago’s Research Computing Center. This work was performed in part at the Aspen Center for Physics, which is supported by National Science Foundation grant PHY-2210452.

\section*{Author Contributions}

MJF, AM, ATS conceived the study, developed methodology, and wrote the original draft. MJF, ATS developed software and performed investigation. MJF, BS, ATS provided formal analysis. MJF, AM supervised and acquired funding. All authors contributed to review and editing. 

\section*{Competing Interests}
The authors declare no competing interests.

\appendix

\section{Kernel-based learning rules that approximate contrastive learning}
\label{app:kernel_constraints}

Our generalized Hebbian learning framework is based on weight updates of the form in Eq. \ref{eqn:w}, replicated here:
\begin{equation}
    \Delta w_{ij} ~ = \epsilon ~ \int_0^{\tau_f + \tau_s} g\left(\int_{-\infty}^{t} K(t-t') s_{ij}(t') \,dt' \right)\, dt ~,
\end{equation}
with $\epsilon$ the learning rate. As discussed in main text Eq. \ref{eqn:saw}, the temporal forcing from the synaptic current $s_{ij}(t)$ is an asymmetric sawtooth wave composed of two linear segments: a fast transition from the free value $s_{ij}^{\rm free}$ to the clamped value $s_{ij}^{\rm clamped}$ occuring over a timescale $\tau_f$, and then a slow relaxation from clamped value $s_{ij}^{\rm clamped}$ back to free value $s_{ij}^{\rm free}$ over a timescale $\tau_s$:
\begin{equation}
s_{ij}(t) = 
 \begin{cases} 
     \frac{A t}{\tau_f} + \overline{s_{ij}} - \frac{A}{2} & t\leq \tau_f \\
      \overline{s_{ij}} + \frac{A}{2} - \frac{A  (t-\tau_f)}{\tau_s} & \tau_f < t\leq \tau_s \\
   \end{cases}
\ ~,
\end{equation}
where $A = s_{ij}^{\rm clamped} - s_{ij}^{\rm free}$ and $\overline{s_{ij}} = \frac{1}{2}(s_{ij}^{\rm clamped} + s_{ij}^{\rm free})$.
The nonlinearity $g$ is also of the form discussed in main text Eq. \ref{eqn:nonlinearity}, i.e. 0 below a threshold magnitude $\theta_g$ and linear elsewhere:
\begin{equation}
g(x) = 
 \begin{cases} 
     x  & \vert x \vert \geq \theta_g \\
      0  & \vert x \vert < \theta_g \\
   \end{cases}~~.
\end{equation}
For the moment, we assume the kernel $K$ to be an arbitrary function with a characteristic timescale $\tau_K$.

In this section, we compute conditions on $K$, $\tau_K$, $\tau_f$, $\tau_s$ and $\theta_g$ such that our weight update Eq. \ref{eqn:w} approximates an ideal contrastive rule:
\begin{equation}
\Delta w_{ij} \approx \epsilon(s_{ij}^{\rm clamped} - s_{ij}^{\rm free}),
\end{equation}
with learning rate $\epsilon$.
We will find that the ideal contrastive rule can be approximated in the following regime:
\begin{enumerate}
\item $I = \int_{0}^{\infty} K(t)dt = 0~$;
\item $\tau_K \ll \tau_f \ll \tau_s~$.
\end{enumerate}

We proceed in three steps. 
First, we will assume that $\tau_K \ll \tau_f$ and show that if $I = 0$, then convolution of linear $s_{ij}$ with kernel $K$ results in an output proportional to $\dot{s}_{ij}$, the time-derivative of $s_{ij}$.
Second, we will show that $\tau_f \ll \tau_s$ in order to distinguish free-to-clamped and clamped-to-free transitions.
Finally, we will empirically show that constrastive updates break down if $\tau_K \rightarrow \tau_f$ from above.


\subsection{Zero-area kernels are required for constrastive updates}

Our first goal is to find conditions for:
\begin{equation}\label{kernel_approx}
    \int_{-\infty}^{t} K(t-t')  s_{ij}(t') \,dt' \approx \frac{ds_{ij}(t)}{dt}.
\end{equation}
This is a prerequisite to reproducing contrastive updates such as Eq. \ref{eqn:wijfreeclamped}.
We will first assume that $\tau_K \ll \tau_f$ and hence that on the timescale of the integral in Eq. \ref{kernel_approx}, $s_{ij}(t) = mt + b$ for some slope $m$ and some y-intercept $b$.

Performing the integration on the LHS of Eq. \ref{kernel_approx} yields:
\begin{equation}
    b \int_{-\infty}^{t} K(t-t') dt' + m \int_{-\infty}^{t} K(t-t') t' dt' .
\end{equation}
By u-substitution ($u = t - t'$) this is:
\begin{equation}
    b \int_{0}^{\infty} K(u) du + m\int_{0}^{\infty}tK(u) du \,-\, m \int_{0}^{\infty}u K(u)du ,
\end{equation}
which results in the following:
\begin{equation}\label{conv}
   \int_{-\infty}^{t}K(t-t') s_{ij}(t')dt'  = (mt + b)I - m(M^{K}_{1}) ,
\end{equation}
where $(M^{K}_{1})$ is the first moment of $K(t)$ around zero and $I$ is the area under the kernel provided $K(t)$ = 0 on $(-\infty, 0)$. Putting this together yields:
\begin{equation}\label{measured_value}
   \int_{-\infty}^{t}K(t-t') s_{ij}(t')dt' = s_{ij}(t)I - m(M^{K}_{1}) ~,
\end{equation}
Here $s_{ij}(t)$ is the instantaneous signal value, $M^{K}_{1}$ is the first moment of $K(x)$ around 0 and $I$ is the total area under the kernel. In order for this integral to equal $m$, it is desirable that:
\begin{equation}
I = 0, ~~ M^{K}_{1} = -1 ~.
\end{equation}

Eq. \ref{measured_value} implies that if we choose kernels $K$ with integrated area $I = 0$, then convolution of $K$ with a linear synaptic current $s_{ij}(t)$ will extract values proportional to the slope $m$ of the synaptic current. 
Further, we can choose the first moment of $K$ such that the value of the convolution is equal to the slope $m$ and not just proportional, though this is not strictly necessary.

\subsection{Separation of slow and fast training timescales is optimal for contrastive updates}

The next step 
is to investigate constraints on the nonlinearity threshold $\theta_g$, $\tau_f$, and $\tau_K$. 

In the previous section, we showed that if $I = 0$ and $\tau_K \ll \tau_f$, then we can approximate the convolution of kernel $K$ with $s_{ij}$ as being equal to $\dot{s}_{ij}$. In this limit, we can rewrite the weight update Eq. \ref{eqn:w} as: 
\begin{equation}
\epsilon^{-1} \Delta w_{ij} ~\approx \int_{0}^{\tau_f} g\left( \frac{ds_{ij}}{dt}\right) dt + \int_{\tau_f}^{\tau_f + \tau_s} g\left(\frac{ds_{ij}}{dt}\right)dt,
\end{equation}
with learning rate $\epsilon$.

In order to reproduce contrastive updates, we require the first integral to not vanish over its entire range of integration, while the second integration should vanish uniformly. This implies that, for a synaptic current of amplitude $A = s_{ij}^{\rm clamped} - s_{ij}^{\rm free}$, fast timescale $\tau_f$, and slow timescale $\tau_s$, the nonlinearity threshold $\theta_g$ must satisfy:
\begin{equation}
    \frac{A}{\tau_s} < \theta_g < \frac{A}{\tau_f} .
\end{equation}

If we require contrastive learning to occur over a range of amplitudes between $A_{min}$ and $A_{max}$ for fxed $\theta_g$, then this condition written over all amplitudes becomes:
\begin{equation}
    \frac{A_{max}}{\tau_s} < \theta_g < \frac{A_{min}}{\tau_f} .
\end{equation}
Saturating this bound yields:
$$\frac{\tau_s}{\tau_f} >  \frac{A_{max}}{A_{min}},$$
implying that our protocol has optimal dynamic range of amplitude when $\tau_f \ll \tau_s$.

\subsection{Kernel timescales must be faster than training timescales for contrastive updates}

In deriving the conditions $I=0$ and $\tau_f \ll \tau_s$ for successful contrastive updates, we assumed $\tau_K \ll \tau_f$.
The latter inequality allowed us to approximate the instantaneous convolution of kernel $K$ with synaptic protocol $s_{ij}$ as convolution with a linear segment, ignoring the exact sawtooth functional form of $s_{ij}$.

\begin{figure}
\includegraphics[width=.95\linewidth]{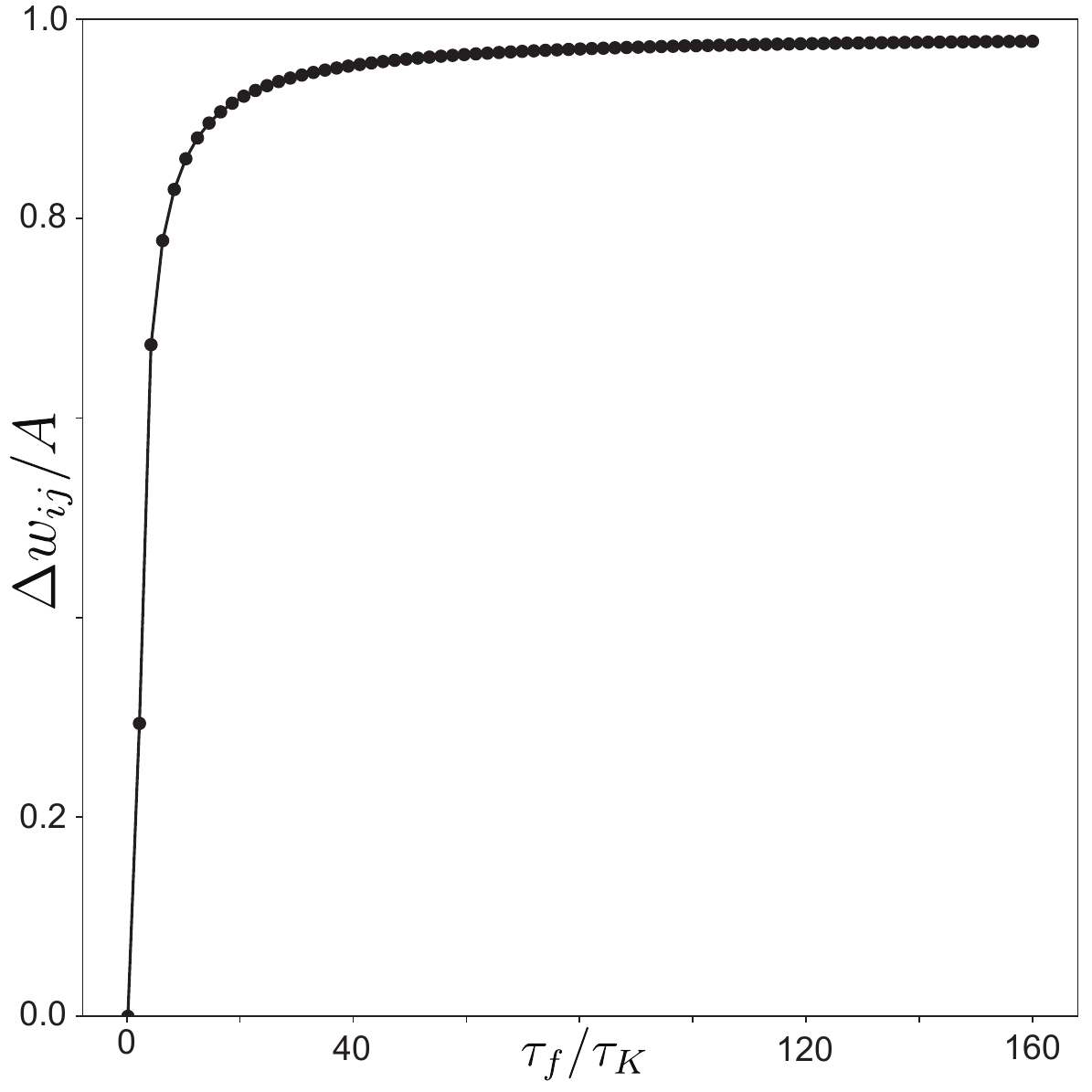}
\caption{\textbf{Finite-time derivative breakdown for short signals.}
Simulation of kernel convolution ability to capture amplitude of a signal with short timescale. x-axis shows ratio between timescale of the signal upswing $\tau_f$ and the memory kernel $\tau_K$. The y-axis gives the ratio of weight update (as given in Eq. \ref{eqn:w}) to the actual amplitude of the signal. A ratio of 1 reflects the kernel convolution approximating well the finite time derivative. At high $\tau_f$, the ratio approaches 1 as desired. At low $\tau_f$, it becomes unreliable.
Convolutions were run numerically for 80 different values of $\tau_f \in (0,8)$, convolving a signal with amplitude 20, with $dt = 1e-4$. Here we fix $\tau_s$ = 20. The kernel used was of the form given by Eq. \ref{eqn:ecoli} and had net 0 area. The nonlinearity threshold $\theta_g$ was chosen to be the average of the lowest $u_{ij}$ value on the upswing and the highest $u_{ij}$ value on the downswing. }
\label{fig:timescales}
\end{figure}

We additionally show that $\tau_K \ll \tau_f$ is required empirically.
For fixed $\tau_K$ and $\tau_s$, we varied $\tau_f$ and computed the weight update computed using Eq. \ref{eqn:w}, using a kernel of a form discussed in Eq. \ref{eqn:ecoli}.
We compared this to the ideal contrastive update, which is equal to the amplitude of the protocol $s_{ij}$.
We find that $\tau_K \ll \tau_f$ in order for the weight update to approach the ideal contrastive update (Fig. \ref{fig:timescales}).

\color{black}
\section{Related Work}
\label{app:related_work}

In this section, we review existing methods to implement contrastive rules, and we show how our proposal compares with these methods.

\paragraph{Storing the states.} 
The contrastive learning rule
\begin{equation}
\Delta w_{ij} = \epsilon \left( s_{ij}^{\rm free} - s_{ij}^{\rm clamped} \right)
\end{equation}
requires a comparison between free and nudge states, but the network cannot be in both states simultaneously. The first option, used in the experiments of \citet{yi2023activity} on memristor crossbar, and \citet{laydevant2024training} on D-wave's Ising machine, is to store the states using an external memory. 

\paragraph{Coupled learning.}
In their experimental realizations of contrastive learning on resistor networks, \citet{dillavou2022demonstration,dillavou2023machine} used two copies of the network, one for each state. 
\citet{wycoff2022desynchronous} demonstrated in a resistor network that weights may be updated asynchronously, but they still require contrasting two states.

\paragraph{Switching between Hebbian and anti-Hebbian updates.}
To avoid storing states or using two networks, another solution consists in performing two updates rather than one - one update for each of the two states (free and nudge).
\begin{enumerate}
\item \textbf{Free phase.}  Allow the network to reach the free state and we perform a `Hebbian' weight update:
\begin{equation}
\Delta w_{ij} = \epsilon s_{ij}^{\rm free}.
\end{equation}
\item \textbf{Clamped phase.} Allow the network to reach the clamped state and we perform a `anti-Hebbian' weight update:
\begin{equation}
\Delta w_{ij} = - \epsilon s_{ij}^{\rm clamped}.
\end{equation}
\end{enumerate}
Another closely related method, proposed by \citet{williams2023flexible} is, for each training example, to perform only one of the two updates, drawn at random. This method provides an unbiased, but high variance, estimator of the weight gradient. \citet{williams2023flexible} also explore schemes where the weight update is activated during the entire phase, and with phases of variable duration (before reaching steady state). However, these methods requires two different weight update mechanisms for each of the two phases: they require switching between Hebbian and anti-Hebbian updates.

\paragraph{Continual EP.}
\citet{ernoult2020equilibrium} rewrite the contrastive learning rule as
\begin{equation}
s_{ij}^{\rm free} - s_{ij}^{\rm clamped} = \int_{\rm free}^{\rm clamped} \frac{ds_{ij}}{dt}dt,
\end{equation}
along any trajectory driving the system from free state to clamped state. Using this formulation of the learning rule, they propose an implementation of contrastive learning with continual weight updates in the clamped phase. The algorithm proceeds as folows:
\begin{enumerate}
\item \textbf{Free phase.} Let the network to reach the free state, while deactivating any weight update, i.e.
\begin{equation}
\label{eq:continual-ep-free}
\frac{dw_{ij}}{dt} = 0.
\end{equation}
\item \textbf{Clamped phase.} Starting from the free state, let the network reach the clamped state, while turning on the weight update
\begin{equation}
\label{eq:continual-ep}
\frac{dw_{ij}}{dt} = \epsilon \frac{ds_{ij}}{dt}
\end{equation}
along the trajectory from free state to clamped state.
\end{enumerate}
In this approach again, one caveat is that it requires switching between two distinct weight update mechanisms. Our algorithm overcomes this hurdle.

\paragraph{Using oscillations.} \citet{baldi1991} proposed to use a periodic nudging signal, e.g. a sinusoidal nudging signal with amplitude $\beta$ and frequency $w$. The weights are then updated using the system's response to this periodic nudging signal, denoted as $s(t)$, as
\begin{equation}
\label{eq:baldi-pineida}
    \Delta w_{ij} = \epsilon \int_0^{2\pi/\omega} \sin(\omega t) s_{ij}(t) dt.
\end{equation}
where $T=2\pi/\omega$ is the period. Similar to other schemes, one caveat of this learning rule is that it requires being modulated by a time-varying factor $\sin(\omega t)$. \citep{anisetti2024frequency} reused the idea of oscillations but, instead of the learning rule of Eq.~\eqref{eq:baldi-pineida}, they showed that in resistor, flow and elastic networks, the weight update for $w_{ij}$ can be performed using the mean and amplitude of $s_{ij}(t)$. However, it remains unclear how the mean and amplitude of the signal can be easily extracted.

\paragraph{Connection with STDP.} 

Spike-timing-dependent plasticity (STDP) (\citet{bi2001synaptic}) can be formulated as a learning rule for asymmetric networks where a synapse distinguishes between the time of presynaptic spikes $t_{k}$ and the time of postsynaptic spikes $t_{j}$:
\begin{equation}
\label{eq:stdp}
\Delta w_{ij} = f(t_{j} - t_{k})
\end{equation}
where $f$ has a positive lobe for positive arguments and a negative lobe for negative arguments, and  approximately vanishes outside a pairing range $[-\tau,\tau]$. 

A large body of work has interrogated the equivalence of spike- and rate-based formulations of STDP\cite{burkitt2004spike,moraitis2020optimality}. 
In particular, \citet{xie1999spike} considered  additive STDP updates for $\delta$-function timeseries $s_j(t)$, written as:
\begin{equation}
\label{eq:xie_kernel}
\Delta w_{ij} = \int_0^T dt_k \int_{-\infty}^{\infty} dt_j ~ f(t_{j} - t_{k}) s_j(t_{j}) s_k(t_{k})
\end{equation}
over a time period $[0, T]$.
They showed that Eq. \ref{eq:xie_kernel} can be approximated as rate-based formulations depending on firing rates $\nu_j(t)$ when rates of spiking vary slowly compared to the pairing range:
\begin{equation}
\label{eq:xie_stdp}
\Delta w_{ij} = \int_0^T dt ~ [\beta_0\nu_j(t) + \beta_1\dot{\nu_j}(t)] \nu_k(t)  
\end{equation}
where $\beta_0$ is the integral of $f$ over the pairing range, and $\beta_1$ is the first moment of $f$ over the pairing range. 

In the case where $\beta_0 = 0$ and $\beta_1 > 0$, Eq. \ref{eq:xie_stdp} can be interpreted as the asymmetric version of the nudge phase update of Continual EP (Eq. \ref{eq:continual-ep}):
\begin{equation}
\label{eq:asym_continual-ep}
\frac{dw_{ij}}{dt} = \beta_1 \dot{\nu_j}(t)\nu_k(t).
\end{equation}

\citet{martin2021eqspike} proposed a spiking version of contrastive learning to train spiking networks with bidirectional weights. Similar to the Continual EP approach, in the free phase, no weight update occurs, and in the nudge phase, weights are updated through spikes based on a learning rule similar to Eq.~\eqref{eq:continual-ep}. They show that a version of STDP emerges from this learning rule.

\paragraph{Competitive Hebbian updates.}

In a directional context with presynaptic neuron $i$ and postsynaptic neuron $j$,
\citet{journe2022hebbian} consider weight updates of the form:
\begin{equation}
\label{eq:softhebb}
\Delta w_{ij} = \epsilon y_j (x_i - u_j w_{ij})
\end{equation}
where $x_i$ is the activation of neuron $i$ and $u_j$ is the weighted input to neuron $j$;
$y_j$ is a winner-take-all factor across all $K$ neurons in a given layer, parameterized by a factor $b$:
\begin{equation}
y_j = \frac{b^{u_k}}{\sum^K_{l=1}b^{u_l}}.
\end{equation}
This method of learning is well-suited to a neuronal context.
However, implementing it in less internally complex physical and biological systems would be a non-trivial task, requiring e.g. the engineering of equivalents to pre- and post-synaptic neurons, as well as the computation of the $y_j$ Boltzmann factor.
The latter could be accomplished potentially through lateral inhibition within layers\cite{krotov2019unsupervised}, at the cost of making architectural assumptions that would narrow the range of natural settings the method could be implemented in.

\paragraph{Agnostic Equilibrium Propagation (AEP).}

\citet{scellier2022agnostic} introduced another dynamical version of EP called AEP, in which the trainable weights ($w$) evolve under physical dynamics to optimize the cost function. In AEP, both the weights ($w$) and the state variables ($\eta$) minimize the system's energy. The weights are strongly coupled to \emph{control variables} ($v$) by a coupling term $U(v,w)$, so the total energy is $U(v,w) + E(w,\eta)$. The control variables can be in either of two states: clamped, or maintain homeostatic control over $w$.

At each training step $t$ with current weights $w^{(t)}$, AEP proceeds as follows. In the free phase, an input is presented, and $w$ and $\eta$ jointly minimize the energy function, while $v^{(t)}$ performs homeostatic control, keeping $w$ at its value $w^{(t)}$, i.e.
\begin{equation}
(w^{(t)},\eta^{\rm free}) = \underset{(w,\eta)}{\arg \min} \left[ U(v^{(t)},w) + E(w,\eta) \right].
\end{equation}
In the clamped phase, the control $v^{(t)}$ is held fixed, and a nudging term $\beta C(\eta)$ is added to the system's energy (where $C$ is the cost function to optimize and $\beta$ is the nudging strength). The system settles to
\begin{equation}
(w^{(t+1)},\eta^{\rm clamped}) = \underset{(w,\eta)}{\arg \min} \left[ U(v^{(t)},w) + E(w,\eta) + \beta C(\eta) \right].
\end{equation}
\citet{scellier2022agnostic} show that the weight change $w^{(t+1)}-w^{(t)}$ between free and clamped phases approximates one step of gradient descent on $C$.

In contrast, our Temporal Contrastive Learning (TCL) approach neither relies on homeostatic control nor requires $w$ to minimize the system's energy.

\paragraph{Temporal Contrastive Learning (our method).}
In contrast to these other methods, we propose a mechanism based on integral feedback that uses a unique weight update rule (a unique learning mode) for both phases. Our proposal is related to, but different from, Continual EP, as we explain next.

The central idea in our approach is to introduce a kernel-based non-equilibrium memory $u_{ij}(t)$ for each variable $s_{ij}$, to approximate the time derivative $\frac{ds_{ij}}{dt}$,
\begin{equation}
    u_{ij}(t) := \int_{-\infty}^t K(t-t') s_{ij}(t') dt' \approx \frac{ds_{ij}}{dt}.
\end{equation}
Intuitively, the memory kernel $K(t-t')$ approximates the derivative of the delta function (Dirac function), in the sense of distribution theory. We then use this non-equilibrium memory to drive the weight updates similar to Eq.~\eqref{eq:continual-ep},
\begin{equation}
    \frac{dw_{ij}}{dt} = g(u_{ij}(t)).
\end{equation}
However, we also need any weight update to vanish on the trajectory from the nudge state to the free state, as in Eq.~\eqref{eq:continual-ep-free}. To achieve this, we introduce two distinct time scales, $\tau_{\rm clamped}$ and $\tau_{\rm free}$ with $\tau_{\rm free} \gg \tau_{\rm clamped}$, so it takes a time $\tau_{\rm clamped}$ to go from free state to nudge state, and a time $\tau_{\rm free}$ to go from nudge state back to free state. Moreover, we use the nonlinear function
\begin{equation}
    g(u) := \begin{cases} 
     u  & \vert u \vert \geq \theta_g \\
      0  & \vert u \vert < \theta_g \\
   \end{cases}~~,
\end{equation}
to suppress inputs below the characteristic threshold $\theta_g$. The network is always in the same learning mode, whether in free or nudge phase. We break the symmetry between free and nudge phase, by using a fast change in the nudge phase (small $\tau_{\rm clamped}$ ) and subsequent slower relaxation in the free phase ($\tau_{\rm free} \gg \tau_{\rm clamped}$), so that the nonlinearity $g$ suppresses weight updates that might occur during the slow nudge-to-free transition.
\color{black}

\color{black}
\section{Quantitative time and energy comparison of TCL vs other contrastive methods}

Our work here on TCL is a proposal for how to implement contrastive learning physically in a fully analog manner; other related proposals (e.g., FF and other CL methods) typically involve digital processing to store two states. Consequently, the primary advantage of our method compared to contrastive algorithms is that it eliminates the need to store and fetch an explicit recording of free and clamped states (or, e.g. positive and negative states in the Forward-Forward algorithm\cite{hinton2022forward}). TCL accomplishes this by embedding the difference between  free and clamped states into a dynamical signal which is generated by integral feedback. This integral feedback mechanism allows for the accurate computation of contrastive weight updates (Eq. \ref{eqn:w}) at the expense of time and energy.

For TCL, the time cost of computing Eq. \ref{eqn:w} is set by the dynamic range to which we want to be able to measure the contrastive difference. For example, if we want to be able to distinguish differences between free and clamped states by a ratio of $10^{-2}$, this would require $\sim 10^3 \tau_K$, where $\tau_K$ is the timescale of the physical processes generating integral feedback in the analog system (Fig. \ref{fig:timescales}). Similarly, there is a cost associated with the energy dissipation rate $\sigma$ of running integral feedback in each synapse, which takes about $\sim 5-10 kT$ in order to achieve $10^{-2}$ relative error to the explicit contrastive update Eq. \ref{eqn:w} (Fig. \ref{fig:markov_cartoon}). The total energy cost of running the network per cycle therefore would be  $\sim 10 kT N_{synapse} (\tau_f + \tau_s)$, with the number of synapses $N_{synapse}$ scaling with the number of nodes $N$ or $N^2$ depending on network connectivity.

For explicitly contrastive implementations, one time cost of computing Eq. \ref{eqn:w} is instead the time of storing each state and retrieving it to perform the difference operation. For memory stored on a digital device locally at each node, this is a constant time, $\tau_{fetch}$. For fully digital implementations (e.g. simulations of EP methods on a computer) this time is extensive in $N$, i.e. $N \tau_{fetch}$. This is in addition to the time cost of relaxation of the physical state of the system in response to alternating between clamping and releasing the output nodes, which we label $\tau_{relax}$. Therefore, to simulate, for example Equilibrium Propagation\cite{scellier2017equilibrium} on a computer would naively take a time $2\tau_{relax} + N\tau_{fetch}$ per cycle. In a twinned circuit device as constructed in Refs. \cite{dillavou2022demonstration,dillavou2024machine}, the time per cycle would be $\tau_{relax} + \tau_{fetch}$.

Energy costs associated with explicit contrastive algorithms arise from the need to continually write and erase the state of the system. Assuming that memory is written at each of the $N$ nodes and is required to an accuracy of $n$ bits, a minimum energy required for contrastive algorithms is $NnkT\ln{2}$ following the Landauer estimate\cite{bennett2003notes}. 

We note that it is hard to generate fair, apples-to-apples comparisons of energy and time costs between TCL and other contrastive algorithms because we do not yet have concrete analog proposals for other contrastive algorithms; some key aspect of those algorithms require digital operations\cite{kendall2020training,dillavou2022demonstration} such as using digital memory to actually compute Eq. \ref{eqn:w}. Nevertheless, we have presented quantitative costs for the TCL proposal to allow for easy comparison to any new analog contrastive learning proposals as they become available. 

\color{black}

\section {Training multi-synapse networks}
\label{app:multi_synapse}

Here, we describe in greater detail how we trained a neural network to classify MNIST digits using our proposed contrastive learning framework. We adapt code from Ref. \cite{scellier2017equilibrium}, originally used to classify MNIST digits using the Equilibrium Propagation algorithm.
We train on a 10000-element subset of MNIST with all digits, with holdout test set of size 2000.
The architecture of the network trained has 784 input nodes, 500 hidden nodes, and 10 output nodes, each with a state $x$.
Each layer is fully connected to the subsequent layer, with no intra-layer connections and no skip connections.

Dynamics of the network during inference minimize the energy functional:
\begin{equation}
E(x) = \frac{1}{2}\Sigma_i x_i^2 - \frac{1}{2}\Sigma_{i, j} w_{ij} \eta_i \eta_{j} - \Sigma_i b_i \eta_i,
\end{equation}
where $b_i$ is the bias of node $i$, $w_{ij}$ is the weight of the synapse connecting nodes $i$ and $j$, and the indices $i, j$ run over the node indices of all layers (input, hidden, and output). Additionally, $\eta$ is a non-linear activation function, in this case $\eta(x) = \text{clip}(x,0,1)$.
During the training of the network, the network instead minimizes:
\begin{equation}
F(x;t) = E(x) + \frac{\beta(t)}{2} \Sigma_{o=0}^{9}(x_{o} - v^{\text{label}}_o)^2 .
\end{equation}
Here, $v^{\text{label}}_o$ is the one-hot encoding of the desired MNIST output class, and the $x_o$ are the states of the 10 neurons in the output layer. $\beta$ is a clamping parameter which varies in time as a sawtooth function.

During both training and inference, the goal is to minimize the energy $F$ with respect to the neuronal state variables $x$. 
A persistent particle trick was employed to keep track of the free phase values for $x$ for a given MNIST entry, to cut down on the amount of time needed to relax to the energy minimum when changing between MNIST entries.
Weights are initialized by the same Glorot-Bengio initialization\cite{glorot2010understanding} as in Ref. \cite{scellier2017equilibrium}, but divided by a factor of 2.
See Ref. \cite{scellier2017equilibrium} Secs. 5.1, 5.2 for further details.

In the original implementation of Ref. \cite{scellier2017equilibrium}, updates to $w_{ij}$ and $b_i$ are proportional to the difference between the state variables in the free state and a single clamped state.
Our goal is instead to subject the neural network to a time-varying sawtooth protocol $\beta(t)$ (e.g. Eq. \ref{eqn:saw}) with a fast upswing from free to clamped states, and a slow relaxation back to free.

In order to approximate this protocol, we move the neural network in small discrete steps from a state with clamping parameter $\beta = 0$ to a state with $\beta = \beta_{max}$, and then back down to the $\beta = 0$ state.
In the numerical experiment reported in Fig. \ref{fig:mnist_example}, we choose $|\beta_{max}| = .5$, with the sign of $\beta_{max}$ chosen randomly, following Ref. \cite{scellier2017equilibrium} Sec. 5.2.
We discretize both the upswing and downswing of the $\beta$ ramp into 9 pieces each, for 20 timepoints total.
We follow the heuristic set in Ref. \cite{scellier2017equilibrium} and allow the system to initially relax for 20 iterations in the free phase, and then for 4 iterations at each subsequent point in the $\beta$ ramp.
Each iteration updates the system following gradient descent of $F$ with respect to the states $x_i$, with a step size of $.5$.
At each point in the ramp, we record the gradient of the energy $F$ with respect to $w_{ij}$ and $b_i$, which are respectively the synaptic current $\eta(x_i)\eta(x_j)$ and the activation $\eta(x_i)$.
This will allow us to compute the weight update Eq. \ref{eqn:w} downstream.

Note that this procedure of needing to record the timeseries of $x$ as $\beta$ is changed is a proxy for the feedback dynamics that we are proposing would be naturally implemented in a physical system.

To map the system behavior onto the timeseries the neural network would experience if forced by an asymmetric sawtooth protocol as in Eq. \ref{eqn:saw}, we use the SciPy interp1d function with default parameters.
We interpolate the values of $x$ onto a timeseries of total time 1, $dt = .01$, and a fast upswing time $\tau_f = .1$ and a slow relaxation time $\tau_s = .9$.

We are now in a position to compute weight updates for our neural network following Eq. \ref{eqn:w}.
We construct a kernel: 
\begin{equation}
\label{eqn:sin_kernel}
    K(t) = \frac{1}{3}\sin(2 \pi t)
\end{equation}
with $0 < t < .1$ and a time resolution $dt = .01$.
The nonlinear threshold function $g$ has a layer-dependent nonlinearity; input-to-hidden layer $w_{ij}$ threshold is $1.2 \times 10^{-6}$, hidden layer $b_i$ threshold is $10^{-5}$, hidden-to-output layer $w_{ij}$ threshold is $2 \times 10^{-5}$, and output layer $b_i$ threshold is $2 \times 10^{-4}$. 
Input layer $b_i$ are not modified during training.

We perform the convolution of our interpolated timeseries of synaptic currents and activations with kernel $K$ (the inner integral of Eq. \ref{eqn:w}) with the SciPy convolve1d function in the ``wrap'' mode for periodic signals.
We divide the result by 100 to account for the time resolution of the signal.
We then pass the convolution through $g$ and integrate the result using the SciPy trapz function with default parameters. 
This result is $\Delta w_{ij}$ from Eq. \ref{eqn:w}.

Next, we perform change existing weights according to $ \frac{10^4 \epsilon}{2 \beta_{max}} \Delta w_{ij}$.
The $\epsilon$ are layer-dependent learning rates, set to .1 for hidden layer associated weights and .05 for output layer associated weights.
For our neural network, we added the factor of $10^4$ empirically to speed convergence.
We perform this procedure of weight updates over minibatches of size 20, with 500 minibatches in the training set and 100 in the testing set.

Finally, we need a method to evaluate the performance of our trained network on predicting MNIST labels for unseen data.
When evaluating the network on the testing set, the network classification of the MNIST image input is the argument of the neuron with the maximal output.
If this argument does not match the label of the MNIST input, then the network has misclassified the input.
In Fig. \ref{fig:mnist_example}B, we report the average classification error after each epoch of training, where a single epoch consists of training the network on all 500 training set minibatches and evaluatng the network on all 100 testing set minibatches.

\section{Kernel coefficients used in assessing speed-accuracy tradeoff}
\label{app:kernel_coefficients}

In order to assess limitations of our proposed contrastive learning framework, we performed a detailed analysis of learning dynamics in a single synapse. 
These results, reported in Fig. \ref{fig:speed_limitations}, made use of kernels that have been studied before in Ref. \cite{celani2010bacterial}.
These kernels have the functional form:
\begin{equation}\label{eqn:ecoli}
K(t-t')~ = ~ \alpha e^{-\tau_K^{-1} (t-t')}((t-t')-\lambda (t-t')^2) .
\end{equation}
Here we briefly detail our choices for the coefficients $\alpha, \tau_K, \lambda$ for such kernels when used in our single-synapse analysis.

First, we fix $\tau_K = \frac{1}{20}$ for quick decay, which simplified numerical calculation.
Second, following Appendix \ref{app:kernel_constraints}, we require $I = \int_{0}^{\infty} K(t)dt = 0$.
Third, again following Appendix \ref{app:kernel_constraints}, our calculations are simplified if the first moment $M^{K}_{1} = -1$.

The latter two requirements allow us to fix $\alpha$ and $\lambda$ as:
\begin{equation}\label{eqn:param_con}
    \alpha = \tau_K^{-3} ~~,\;
    \lambda = \frac{1}{2\tau_K}.
\end{equation}
Numerically, the memory kernel is created with unit length and timestep $dt = 1e-4$.

\section{Weight updates and optimal thresholds for kernels with non-zero area}
\label{app:AUC_kernels}

Our single synapse analysis was first performed in an ideal limit, assuming kernels with zero area.
In Fig. \ref{fig:nonzero_area_kernels}, we generalized those results to non-ideal scenarios where kernels have non-zero integrated area.

In this appendix we discuss how weight updates are modified when we leave the integrated area $I = \int_{0}^{\infty} K(t)dt = 0$ regime, which is ideal for contrastive updates (see Appendix \ref{app:kernel_constraints} for the $I = 0$ case).
We also address the question of choosing the threshold value $\theta_g$ for the nonlinearity $g$ in the $I \neq 0$ regime.
We provide a brief note on constructing kernels with a controlled amount of integrated area $I$.

\subsection{Weight updates for kernels with non-zero area}
In Eq. \ref{measured_value}, we derived the relation:
\begin{equation}
     \int_{-\infty}^{t} K(t-t') s_{ij}(t') \,dt' = s_{ij}(t)I - m(M^{K}_{1}),
\end{equation}
in the limit as $\tau_K \ll \tau_f$ for an arbitarry kernel $K$ and a linear $s_{ij}(t) = mt+b$. $M^{K}_{1}$ is the first moment of the kernel.

Evaluated on the fast rise of time $\tau_f$ from free to clamped states, $m = \frac{(s_{ij}^{\rm clamped} - s_{ij}^{\rm free})}{\tau_f}$. If we assume $M_0^K = -1$, but $I \neq 0$, we have the following:
\begin{equation}
\int_{-\infty}^{t}K(t-t') s_{ij}(t')dt' \approx \frac{(s_{ij}^{\rm clamped} - s_{ij}^{\rm free})}{\tau_f} + Is_{ij}(t).
\end{equation}
Then integrated over a cycle (assuming we can always threshold the slow relaxation from clamped to free out), this yields for the weight update:
\begin{equation}
\int_0^{\tau_f} \int_{-\infty}^{t}K(t-t') s_{ij}(t')dt' dt = (s_{ij}^{\rm clamped} - \,s_{ij}^{\rm free}) + I\tau_f\overline{s_{ij}},
\end{equation}
where $\overline{s_{ij}} = (s_{ij}^{\rm clamped} + s_{ij}^{\rm free})/2$. Thus an offset of $I\tau_f \overline{s_{ij}}$ is added to the desired weight update (Fig. \ref{fig:nonzero_area_kernels}B).

\subsection{Choosing optimal thresholds for kernels non-zero area}
Next, we discuss an optimal choice for $\theta_g$ when $I \neq 0$. 

The sawtooth wave used as the signal here takes to following form, as replicated from Eq. \ref{eqn:saw}:
\begin{equation}
s_{ij}(t) = 
 \begin{cases} 
     \frac{A t}{\tau_f} + \overline{s_{ij}} - \frac{A}{2} & t\leq \tau_f \\
      \overline{s_{ij}} + \frac{A}{2} - \frac{A  (t-\tau_f)}{\tau_s} & \tau_f\leq t\leq \tau_s \\
   \end{cases}
\ ~~.
\end{equation}
The value of the convolution measured in Eq. \ref{measured_value} takes on two distinct forms on the two sections of the sawtooth wave. Plugging $s_{ij}(t)$ from Eq. \ref{eqn:saw} into Eq. \ref{measured_value} 
we have:
\begin{equation}\label{eqn:d2}
u_{ij}(t) = 
 \begin{cases} 
      \frac{A}{\tau_f} + I  (\frac{A t}{\tau_f} + \overline{s_{ij}} - \frac{A}{2}) & t\leq \tau_f \\
      - \frac{A}{\tau_s} + I  (\overline{s_{ij}} + \frac{A}{2} - \frac{A  (t-\tau_f)}{\tau_s}) & \tau_f\leq t\leq \tau_s \\
   \end{cases}.
\end{equation}

Since the performance of a kernel relies on its ability to capture the difference between the two sections of the sawtooth signal, one sufficient condition for a good protocol is the following: 
\begin{equation}\label{suff_cond}
\min\left(\int_{-\infty}^{t}K(t-t') s_{ij}(t')dt' ~ \text{on} ~ (0,\tau_f)\right) > \theta_g  \end{equation}
\begin{equation}  
\theta_g > \max\left(\int_{-\infty}^{t}K(t-t') s_{ij}(t')dt' ~ \text{on} ~ (\tau_f,\tau_s)\right) .
\end{equation}

This condition ensures the values that are desired are all above threshold, while the values that should be suppressed (those on the slow relaxation timescale), are all thresholded out. It is a stronger condition than may be absolutely necessary for acceptable performance. This condition enforced onto $\int_{-\infty}^{t}K(t-t')s_{ij}(t')dt'$ as represented in Eq. \ref{eqn:d2} implies:
\begin{equation}\label{ineq}
\frac{A}{\tau_f} + I  (\overline{s_{ij}} - \frac{A}{2}) > - \frac{A}{\tau_s} + I  (\overline{s_{ij}} + \frac{A}{2}) ~, 
\end{equation}
assuming both sides are positive. This assumption, in the limit as $A \ll \overline{s_{ij}}$ requires $I$ and $\overline{s_{ij}}$ to be the same sign. 

In the limit as $A \gg \overline{s_{ij}}$, it requires:
\begin{equation}
\frac{2}{\tau_s}<I<\frac{2}{\tau_f} .
\end{equation}
The condition itself implies:
\begin{equation}
I < \frac{1}{\tau_f} + \frac{1}{\tau_s} .
\end{equation}
Combined, this implies:
\begin{equation}\label{eqn:Ilim}
\frac{2}{\tau_s}< I < \frac{1}{\tau_f} + \frac{1}{\tau_s} .
\end{equation}
This defines the tradeoff between timescales and kernel area that, if satisfied, guarantees the existence of a $\theta_g$ which distinguishes between the two sections of the sawtooth. In this work, we choose $\theta_g$ as the maximum of the second portion of convolution values:
\begin{equation}\label{eqn:theta}
\theta_g = - \frac{A_{max}}{\tau_s} + I  (\overline{s_{ij}} + \frac{A_{max}}{2}) .
\end{equation}
Per Eq. \ref{eqn:Ilim}, this choice of $\theta_g$ is valid in regimes where $I$ is sufficiently large to make the term involving I in Eq. \ref{eqn:theta} dominate. This ensures that the expression \ref{eqn:theta} is the absolute value of the most extreme value of $u_{ij}$ on the downswing (i.e. the downswing is entirely thresholded out by $g$). This ``adaptive'' choice of $\theta_g$ is tuned to the maximum signal amplitude within an otherwise fixed protocol. It guarantees that the downswing of the sawtooth is thresholded out for all values of $A < A_{max}$, while for some values, portions of the upswing are above threshold. The inequality in Eq. \ref{ineq} guarantees that the entire upswing is above threshold when $A = A_{max}$. However, this choice of $\theta_g$ leads to good performance even when inequality \ref{ineq} isn't strictly met. This choice of $\theta_g$ is used in our simulations whenever $I$ is non-negligible.


\subsection{Coefficients for kernels with non-zero area}
Here, we provide a brief note for how to construct kernels used in the Fig. \ref{fig:nonzero_area_kernels} analysis, which is done for kernels with $I \neq 0$.
In particular, the functional form of the kernels we chose is given by Eq. \ref{eqn:ecoli} but with $I \neq 0$.
The following parameter constraints provide the desired integrated area $I$ while maintaining the first moment $M^1_K = -1$:
\begin{equation}
K(t-t')~ = ~ \alpha e^{-\tau_K^{-1} (t-t')}((t-t')-\lambda (t-t')^2)
\end{equation}
\begin{equation}
    \alpha = 3I\tau_K^{-2} + \tau_K^{-3}
\end{equation}
\begin{equation}
    \lambda = \frac{\tau_K^{-2} + 2I\tau_K^{-1}}{6I + 2\tau_K^{-1}}.
\end{equation}
These parameter choices reduce to those given in Eq. \ref{eqn:param_con} when $I = 0$. 

\section{Protocol evaluation metrics}

Below we outline two metrics that we used to evaluate the performance of our proposed learning protocol across Figs. \ref{fig:speed_limitations}-\ref{fig:markov_cartoon}.

Both metrics are derived from a plot of the computed weight update value $\Delta w_{ij}$ and the sawtooth amplitude $A$ as shown in Fig. 4B; a ``weight update curve." This graph is created by calculating $\Delta w_{ij}$ for signals over a range of amplitudes and recording pairs $(A, \Delta w_{ij})$ for 100 values of $A$ between 0 and $A_{max}$ using the linspace function from NumPy. 

There are two main features that we extract from the weight update curve: offset and dynamic range. 
To calculate these values in the limit of low $I$, we use the PiecewiseLinFit function from the pwlf library to section the response curve into three linear segments. The offset is calculated as the y-intercept of the extension of the rightmost linear segment using the intercepts function from the same library. If the rightmost linear segment has a slope between .75 and 1.25 (a proxy for approximately a linear response with slope = 1) calculated with the slopes function from the same library, then $A_{min}$ is defined as the left edge of this segment, calculated using the breaks function from pwlf. 

This formulation is motivated by the observation that generated response curves had three regions. At low A, there is no response, then response spikes sharply as the threshold is passed, and then approximately linearly grows. The left edge of this segment therefore is a proxy for the range of approximately linear response. The offset is a proxy for the impact on background signal value which shifts this segment up or down. We use dynamic range = $\frac{A_{max}}{A_{min}}$ as the most explanatory measure. 

\section{Kernels based on Markov state models}
\label{app:markov_chain}

In addition to investigating the effect of memory kernels described at a phenomenological level, we also investigated kernels derived from microscopic statistical mechanics models.
Building off work by Ref. \cite{lan2012energy}, who constructed a Markov state model for \textit{E. coli} chemoreceptor adaptation, we used the model to construct kernels and investigated how varying the parameters in the network affected the kernels produced.
These kernels allowed us to interrogate the relationship between dissipation and performance in our proposed contrastive learning framework (Fig. \ref{fig:markov_cartoon}).

First we create a Markvo chain with 10 nodes, as in Ref. \cite{lan2012energy}. The Markov chain is laid out as a 2 x 5 grid where nodes are connected only to their neighbors on the square grid, resembling a ladder. The occupancies of each node as well as the transition rates between nodes are stored in lists.

\begin{figure}
\includegraphics[width=.95\linewidth]{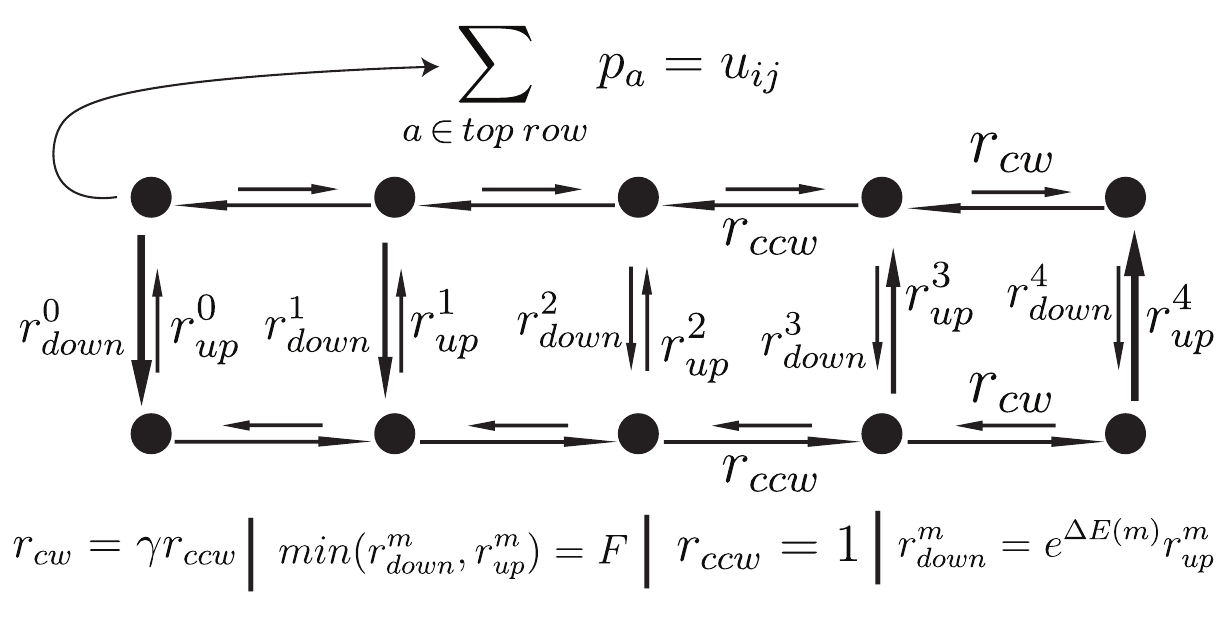}
\caption{\textbf{Structure of kernel-generating Markov state model.}
The adaptive dynamics within each synapse are modeled with a 10-state Markov state model laid out as a 2x5 grid. The horizontal rates ($r_{cw}, r_{ccw}$) are parameterized by $\gamma$. The vertical rates $(r^m_{up}, r^m_{down})$ are parametrized by F and $\Delta E(s_{ij},m)$. The sum of top row occupancies $p_a$ is identified with $u_{ij} = \int_{-\infty}^{t} K(t-t')  s_{ij} \, dt'$. The time derivative of $u_{ij}$ then defines the memory kernel: $\frac{d}{dt}u_{ij}(t) = K(t)$.
}
\label{fig:markov_chain}
\end{figure}

The evolution of a Markov chain is governed by the master equation given as:
\begin{equation}\label{master}
\frac{d\vec{p}}{dt} = \boldsymbol{R}\vec{p} ~.
\end{equation}
Here $\vec{p}$ is a vector containing the occupancies of each node, while $\boldsymbol{R}$ is the transition rate matrix, and $\boldsymbol{r}_{ij}$ is the transition rate from node j to node i. Due to the connectivity of the network shown in Fig. \ref{fig:markov_chain}, most of the rates are 0 and the rest are defined by a small number of parameters and relations given below. The horizontal rates in the direction of clockwise circulation (rightward on the top level and leftward on the bottom), are modulated by a parameter $\gamma$ which controls the dissipation required for non-equilibrium circular flow:
\begin{equation}\label{gamma}
r_{clockwise} = \gamma  r_{counterclockwise} ~ .
\end{equation}
Clockwise rates are fixed to the value 1 in this work for simplicity. There are 5 sets of vertical rates, $r^{n}_{up}$ and $r^{n}_{down}$ where $m$ is an integer between 0 and 4 representing the horizontal position. There is an energy difference $\Delta E$ defined for each $m$ given as:
\begin{equation}\label{E}
\Delta E = 2(1-m) + \ln(\frac{1 + (\frac{s_{ij}}{18.2})}{1 + (\frac{s_{ij}}{3000})}) ~.
\end{equation}
Here $s_{ij}$ is the instantaneous signal value. Following the model in Ref. \cite{lan2012energy}, the vertical rates are a function of $\Delta E$:
\begin{equation}\label{ws}
r^{m}_{down} = ~e^{\Delta E}  r^{m}_{up} ~ ~ ~ ~, ~~ ~ min(r^{m}_{down}, r^{m}_{up}) ~ = ~ F,
\end{equation}
where F is a constant that governs the ratio between vertical and horizontal transition rates. It is default set to F = 25. A transition matrix can be constructed for any choice of rates defined above.

The process of finding the memory kernel requires bringing the Markov chain to a steady state and then perturbing it. The null\_space function from the SciPy library linalg is used to get the null space of the transition matrix, which is the steady state of the Markov chain. A short time-series of occupancy values are captured for the Markov chain sitting in the steady state. 

Time evolution of the Markov chain uses the odeint function from the SciPy integrate library to get a time-series of occupancy values. The argument ``derivfun" is set to a function that implements the dynamical rule in Eq. \ref{master}. The argument ``y0" is set to the null-space of the Markov chain. 

Starting in the steady state, the transition matrix of the Markov chain is then changed by perturbing the variable $s$ from s = 0 to some higher value in a step function form. This causes flow of occupancy in the Markov chain before settling into a new steady state. As this evolution occurs, the occupancy $u_{ij}$ of the top level of the Markov chain is summed and saved as a time-series of length 5 with density $dt = 5e-4$ for a total of 10,000 values stored in a 1-dimensional array. 

The kernel is then defined as the derivative of the step function response (the graph of top level occupancy over time after step function perturbation of $s_{ij}$). The top level occupancy is identified with the signal $u_{ij}$: \begin{equation}
\sum\displaylimits_{a \,\in\, top \: row}p_a \equiv u_{ij}.
\end{equation}
Then $u_{ij}$ is defined as usual:
\begin{equation}\label{eqn:defu}
u_{ij}(t) = \int_{-\infty}^{t}K(t-t') s_{ij}(t')dt'
\end{equation}
 We use the following general result:
\begin{equation}
\frac{d}{dx}\left(\int_{-\infty}^{x}f(x-t') g(t')dt' \right)= \int_{-\infty}^{x}f(x-t') \frac{dg}{dx}(t')dt'
\end{equation}
We substitute into Eq. \ref{eqn:defu}.
\begin{equation}
\frac{d}{dt}u_{ij}(t) = \int_{-\infty}^{t}K(t-t')\frac{ds_{ij}}{dt}(t')dt'
\end{equation}
For $s_{ij}$ chosen to be a step function $S$, we have
\begin{equation}
\int_{-\infty}^{t}K(t-t')\frac{dS}{dt}(t')dt' = \int_{-\infty}^{t}K(t-t')\lambda(t-t')dt' = K(t) . 
\end{equation}
We thus arrive at our process of extracting the kernel $K$ from each Markov chain: 
\begin{equation}
\frac{d}{dt} \left( \sum\displaylimits_{a \,\in\, top \: row}p_a\right) = K(t)
\end{equation}
The derivative function from the SciPy misc library is used to calculate this derivative, passing ``order = 3". Each kernel $K$ is then normalized to have first moment $M^K_1$= -1. 

One other relevant measure that is associated with a Markov chain, is its dissipation, defined as:
\begin{equation}\label{SIdiss}
\sigma = \sum_{i > j}({r}_{ij}p_j - {r}_{ji}p_i)\ln(\frac{{r}_{ij}p_j}{{r}_{ij}p_j}) ~. 
\end{equation}
This quantity is calculated in the steady state, and over time during the evolution of the Markov chain. 

\section {Implementation details for Figs. \ref{fig:speed_limitations}-\ref{fig:markov_cartoon}}\label{app:plot_details}
\subsection{I = 0 Data}

The data for Fig. \ref{fig:speed_limitations}C was generated in a parameter regime with $I = \overline{s_{ij}} = 0$. NumPy.random.randint was used to randomly choose $\tau_f$ in the range (0.1, 5) and $\tau_s$ in the range (0.2, 50). The \textit{E. coli} kernel is constructed with specified area $I = 0$. Then the weight update curve is generated with the inputs $A_{max} = 0, \overline{s_{ij}} = 0$ and dynamic range and offset are calculated from that graph. Then $\tau_f$, $\tau_s$, $A_{min}$, and offset are saved into an array. 
We run 100 parallel batches of this calculation, with each batch generating 10 data points for a total of 1000 points of data. All colorbars used in this work are created with the Matplotlib colors library.

\subsection{I = 0.05, $\overline{s_{ij}} = 25$ Data}

The data for Fig. \ref{fig:nonzero_area_kernels}C is generated by largely the same process as that for Fig. \ref{fig:speed_limitations}C, but with some changes in parameters. The average signal $\overline{s_{ij}}$ is set to 25. The area of the kernel $I$ is set to 0.05. 
In the creation of Fig. \ref{fig:nonzero_area_kernels}C, the signal timescale values are converted into units of kernel timescale by multiplying by a factor of $\frac{1}{\tau _k} = 20$, and $A_{min}$ is replaced by dynamic range $\frac{A_{max}}{A_{min}}$ for $A_{max} = 100$. Outliers in the data are removed (points with negative offset, or offset above 0.5), which reflect response curves not of the form analyzable by our offset calculating function, and therefore not suited to approximating contrastive learning with the present scheme. The colorbar is normalized to span the range of values of $\log_{10}(\frac{\tau_f}{\tau _k})$. A 1000 point scatterplot is made for offset vs $\tau_f$.

\subsection{Markov kernel data}
The final batch of data, used for Fig. \ref{fig:markov_cartoon}B, randomly generates kernels using Markov chains and tests them on an otherwise fixed protocol. A total of 5,000 random kernels are generated. The tuneable parameters in the creation of each kernel are F, $\gamma$, and $\Delta s_{ij}$. $F$ and $\gamma$ are defined in Appendix \ref{app:markov_chain}, and $\Delta s_{ij}$ is the height of the step function passed as signal. For every 50 kernels, random values of gamma (10 of order unity and 40 ranging from order 1 to order $10^{-25}$) are generated, and $\Delta s_{ij}$ is chosen randomly from (5, 250) with the randint function from NumPy random library. The rate parameter F is chosen randomly from (1, 100). Each kernel is generated following the process outlined in Appendix \ref{app:markov_chain}. Once the kernels are created, they are normalized to have first moment 1 and their areas are computed. 

For each kernel,  a weight update curve is generated with the parameters $(A_{max} = 50, \tau_f = 3, \tau_s = 15, \overline{s_{ij}} = 10)$ and $\theta_g$ is chosen using the rule of Eq. \ref{eqn:theta}. The kernel, its randomly generated values of $\gamma$, $\Delta s$, $F$, its area, and the lists of x and y values that make up the response curve are all saved into a NumPy array. 

Since the randomly generated Markov kernels have much less well defined and consistent shapes than those generated with low area \textit{E. coli} kernels (the new kernels have area ranging from 0 to order 1), we use a different method of calculating offset and dynamic range. Rather than being distinctly split into three mostly linear regions, the resulting response curves are much more curved. To address this, we defined $A_{min}$ as the minimum amplitude value for which the response curve crossed a threshold value of 0.05. Since the response curves are less linear than those of the low area \textit{E. coli} curves, the offset was redefined here to be the y-intercept of the line of best fit for the data points between $A_{min}$ and $A_{max}$. Next dissipation is calculated using Eq. \ref{SIdiss}. Then the inverse of dynamic range, offset and dissipation for each protocol are saved to a NumPy array of length 5000. 

\end{document}